\documentclass[twocolumn,superscriptaddress,nofootinbib,aps,preprintnumbers,prd]{revtex4-1}

\usepackage{amsmath,amssymb}
\usepackage{graphicx}
\usepackage{color}
\usepackage{hyperref}
\usepackage{url}
\usepackage{slashed}
\usepackage{subfigure}
\usepackage{slashed,bbm}
\usepackage{graphics,psfrag,epsfig}
\usepackage{dsfont}
\usepackage{enumerate}
\usepackage{mathtools}

\newcommand{\be}{\begin{equation}}
\newcommand{\ee}{\end{equation}}
\newcommand{\bee}{\begin{eqnarray}}
\newcommand{\eee}{\end{eqnarray}}

\begin{document}

\title{Critical $O(N)$ models above four dimensions: Small-$N$ solutions and stability}

\author{Astrid Eichhorn}
\email[]{a.eichhorn@imperial.ac.uk} 
\affiliation{Blackett Laboratory, Imperial College, London SW7 2AZ, United Kingdom}
\preprint{Imperial/TP/2016/AE/02}

\author{Lukas Janssen}
\email[]{lukas.janssen@tu-dresden.de} 
\affiliation{Institut f\"ur Theoretische Physik, Technische Universit\"at Dresden, 01062 Dresden, Germany}

\author{Michael M. Scherer}
\email[]{scherer@thphys.uni-heidelberg.de} 
\affiliation{Department of Physics, Simon Fraser University, Burnaby, British Columbia, Canada V5A 1S6}
\affiliation{Institut f\"ur Theoretische Physik, Universit\"at Heidelberg, Philosophenweg 16, 69120 Heidelberg, Germany}

\begin{abstract}
We explore $O(N)$ models in dimensions $4<d<6$. Specifically, we investigate models of an $O(N)$ vector field coupled to an additional scalar field via a cubic interaction.
Recent results in $d=6-\epsilon$ have uncovered an interacting ultraviolet fixed point of the renormalization group~(RG) if the number~$N$ of components of the vector field is large enough, suggesting that these models are asymptotically safe.
We set up a functional RG analysis of these systems to 
address three key issues: Firstly,  we find that in $d=5$ the interacting fixed point exists all the way down to $N=1$. Secondly, 
we show that the standard $O(N)$ universality classes are actually embedded in those of the cubic models, in that the latter exhibit the same values for (most of) the critical exponents, but feature an additional third RG relevant direction.
Thirdly, we address the critical question of global stability of the fixed-point potential to test whether the fixed point can underly a viable quantum field theory.
\end{abstract}

\maketitle

\section{Introduction: $O(N)$ models within the asymptotic-safety paradigm}

It has recently been suggested that a
unitary UV completion of $O(N)$ models \emph{above} four dimensions can be found if the theory is formulated in terms of an $O(N)$ vector field coupled to an additional scalar field, defined by the Lagrangian~\cite{Fei:2014yja}
\begin{align}\label{eq:lagrangian01}
	\mathcal{L}=\frac{1}{2}(\partial_\mu \phi_i)^2+\frac{1}{2}(\partial_\mu z)^2+\frac{1}{2}g z \phi_i \phi_i + \frac{1}{6}\lambda z^3\,,
\end{align}
with $i=1,\dots,N$.
The theory without the kinetic term for $z$ arises from a Hubbard-Stratonovich transformation of a pure quartic $O(N)$ model with $N$-component vector $\phi_i$. The new field $z$ is introduced as a composite field 
representing the quadratic operator $\phi_i^2$.
In $6-\epsilon$ dimensions a perturbative renormalization group (RG) study has uncovered an UV fixed point with real values of the couplings for large enough $N$~\cite{Fei:2014yja, Fei:2014xta}. This provides a possible example of an asymptotically safe model.  Asymptotic safety, proposed by Weinberg \cite{Weinberg:1980gg}, has recently been explored as a new paradigm for quantum field theories~\cite{Gies:2013pma,Litim:2014uca}, including quantum gravity \cite{Reuter:1996cp,Codello:2008vh,Benedetti:2009rx,Falls:2013bv,Becker:2014qya,Dona:2013qba}. 
At its heart lies an interacting fixed point, at which quantum fluctuations render the theory scale invariant.
Further, the model defined by Eq.~\eqref{eq:lagrangian01} is conjectured to be related to Vasiliev higher-spin theories \cite{Fradkin:1987ks,Vasiliev:1990en,Vasiliev:1992av,Vasiliev:1995dn,Vasiliev:1999ba,Vasiliev:2003ev} via the AdS/CFT correspondence \cite{Klebanov:2002ja,Giombi:2014iua}.
It may thus 
provide
a concrete connection between AdS/CFT and asymptotic safety.

Moreover, the model could provide a possible starting point to render the Higgs-Yukawa sector of the Standard Model asymptotically safe: While the issue of triviality represents an important unresolved problem in four dimensions~\cite{Gies:2009hq,Gies:2009sv,Scherer:2009wu,Vacca:2015nta}, an UV fixed point in a setting with an additional compact dimension 
could provide a possible mechanism for UV completion.

Within the $\epsilon$ expansion, a stable%
\footnote{Here, ``stable'' refers to the number of RG relevant directions, not to the stability of the fixed-point potential.}
and real fixed point has been found  only if $N>N_c$, with the critical value being $N_c \approx 1038$ at one loop and $N_c(d=5) \approx 64$ on the level of the three-loop expansion~\cite{Fei:2014xta}. 
Using resummation techniques, a value of $N_c \approx 400$ has been found on the four-loop level~\cite{Gracey:2015tta}. Conformal bootstrap techniques that are applicable to critical $O(N)$ models \cite{Nakayama:2014yia} have also been employed to conjecture the existence of a finite value of $N_c$ in $d=5$ \cite{Chester:2014gqa}, while Ref.~\cite{Bae:2014hia} finds indications for $N_c=1$. In addition, a related model based on the tensorial instead of the scalar Hubbard-Stratonovich decoupling features an $O(N)$ fixed point in $d=6-\epsilon$ for small values $N=2,3$~\cite{Herbut:2015zqa}. 
The situation appears to be similar to the fixed-point structure of the three-dimensional Abelian Higgs model
\cite{coleman1973, halperin1974, kolnberger1990}, where the lowest-order $\epsilon$ expansion significantly overestimates the critical $N$, and alternative methods now agree on a low value~\cite{dasgupta1981,Bergerhoff:1995zm,bergerhoff1996,herbut1996,herbut1997,Freire:2000sx, herbut2007}.

If existent, universality suggests that a stable interacting fixed point in the theory \eqref{eq:lagrangian01} may be equivalent to the usual Wilson-Fisher fixed point of the pure quartic $O(N)$ models above four dimensions. In this latter formulation it is located at negative coupling, and therefore usually rejected as unphysical.
The possible existence of such a duality triggers a number of intriguing questions:
\begin{enumerate}
	\item What is the true $N_c$ in five dimensions?
	\item Does the classical equivalence of the cubic theory defined by Eq.~\eqref{eq:lagrangian01} and the original quartic $O(N)$ model continue on the quantum level?
	\item Does the cubic model feature a stable fixed-point potential?
\end{enumerate}
If the answers to Questions 2 and 3 turned out positive, and the value for $N_c$ in $d=5$ (Question 1) happened to be not too large, the existence of the fixed point in the cubic model would cure the deficiencies of the Wilson-Fisher fixed point in the original quartic formulation of the model, and its rejection would turn out to be premature.
To address these questions, we employ
a nonperturbative approach that
goes beyond the $\epsilon$ expansion,
the functional renormalization group~\cite{Berges:2000ew}. 
A first functional RG analysis in this spirit has been undertaken in \cite{Percacci:2014tfa}. Working within a local potential approximation of the original pure $O(N)$ model with quartic self-interaction (i.e., without parametrization in terms of a Hubbard-Stratonovich field), this early study comes to the conclusion that no  physically admissible fixed-point solution with stable potential exists in $4<d<6$. The formulation within the original quartic model, however, might miss important nonperturbative information that is encoded in the Hubbard-Stratonovich parametrization. Physically, this would imply that momentum-dependent interaction channels would actually be important to recover an admissible fixed point. These can be efficiently captured by using a Hubbard-Stratonovich transformation.
On the other hand, it may also be conceivable that the pure quartic $O(N)$ models and the $O(N)$-symmetric models that include an additional scalar field $z$ turn out to actually \emph{not} be equivalent on the quantum level. Then, a fixed point with stable potential could exist in the latter, while no viable fixed point would exist in the former, i.e., the extension of the Wilson-Fisher fixed point to $d>4$ would indeed feature an unstable potential, as expected from the perturbative expansion.

In the present work, we apply the functional renormalization group to the cubic Lagrangian formulated in Eq.~\eqref{eq:lagrangian01} as an attempt to address all three questions.
The work is organized as follows: In Sec.~\ref{sec:frg} we introduce our method and present the flow equations. The connection to the previous $\epsilon$-expansion results is established in Sec.~\ref{sec:eps-expansion}, while we discuss our results on $N_c$ in Sec.~\ref{sec:critical-N} (Question 1). In Sec.~\ref{sec:universality-classes} we compare the universality class defined by our fixed point with the original $O(N)$ universality classes (Question 2). The discussion of the stability of the fixed-point potential (Question 3) is made in Secs.~\ref{sec:stability-large-N} and \ref{sec:stability-large-N}. We summarize and conclude in Sec.~\ref{sec:conclusions}.

\section{Functional renormalization group}
\label{sec:frg}

The functional renormalization group (FRG) approach is based on a functional RG equation -- the Wetterich equation \cite{Wetterich:1992yh}. 
It allows us to devise truncational schemes to systematically evaluate the scale-dependence of quantum and statistical field theories within and beyond the realm of perturbation theory.

\subsection{Key features}

The Wetterich equation has a one-loop structure, resembling the one-loop functional determinant. This implies that the proliferation of diagrams that arises beyond leading order in the $\epsilon$ expansion is avoided with this method. At the same time, the use of the non-perturbative propagator encodes higher-order effects within the one-loop structure.
Thus, the FRG is considered an ideal tool to investigate interacting RG fixed points, in particular those underlying asymptotically safe models.
As an intriguing example, it provides a way to investigate asymptotically safe quantum gravity away from the critical dimension, which in gravity is $d=2$. 
Employing the $\epsilon$ expansion, early studies have found an interacting fixed point in quantum gravity, providing an UV complete quantum field theory for the metric \cite{epsilongravity}.
As a major success of the functional renormalization group, this fixed point has been found to persist for $d>2$, and in particular at $d=4$, in a variety of approximations~\cite{Reuter:1996cp,Codello:2008vh,Benedetti:2009rx,Falls:2013bv,Becker:2014qya,Dona:2013qba}. 
In this work we explore the interacting fixed point found for the cubic scalar model in $d=6-\epsilon$ in Refs.~\cite{Fei:2014yja, Fei:2014xta} in a similar spirit, starting in the vicinity of $d=6$, in order to be able to compare with the previous results. 
Ultimately, we move away from the critical dimension in order to estimate the fate of the fixed point towards the physical case $d=5$.

\subsection{The Wetterich equation}

The FRG provides a practical implementation of Wilson's idea of successively integrating out degrees of freedom in the functional integral representation.
It is formulated in terms of an exact functional differential equation describing the evolution of the generating functional for the one-particle irreducible correlation functions, i.e., the effective action $\Gamma$, equipped with an IR momentum cutoff scale $k$,  \cite{Wetterich:1992yh},
\begin{align}\label{eqn:Wetterich}
\partial_t\Gamma_k = \frac{1}{2}\text{STr}\left[(\Gamma_k^{(2)}+R_k)^{-1}\partial_t R_k\right],
\end{align}
with $\partial_t=k\partial_k$, see also \cite{Morris:1993qb, Ellwanger:1993mw}.
The scale-dependent effective action $\Gamma_k$ interpolates between the microscopic action $S$ and the full quantum effective action  $\Gamma_{k\rightarrow0}$ in the IR.
The interpolation is implemented by the regulator function $R_k(p)$ which depends on the IR cutoff scale $k$ and the momentum $p$ of the field configurations that are integrated over in the generating functional. 
It suppresses low-momentum fluctuations, as $R_k(p)>0$ for $p^2<k^2$.  
The regulator function satisfies  $R_k(p)\rightarrow \infty$ for $k\rightarrow \Lambda \rightarrow \infty$ and $R_k(p)\rightarrow 0$ for $k/|p|\rightarrow 0$.
The regulator modifies the microscopic action in the functional integral $Z=\int_\Lambda \mathcal{D}\Phi\, e^{-S[\Phi]}$ by the replacement
\begin{align}
S\rightarrow S &+ \int_p\frac{1}{2}\Phi(-p)R_{k}(p)\Phi(p)\,,
\end{align}
where $\Phi$ represents a collective field variable for all field degrees of freedom of a specific model, $\Phi=(\phi_i,z)$ in our case. 
The scale-dependent action is then defined as the (modified) Legendre transform of the regularized Schwinger functional $W_k[J]=\ln Z_k$, see \cite{Berges:2000ew} for details. 
Thus $\Gamma_k$ contains contributions from high-momentum quantum fluctuations at $p^2/k^2>1$, only.
Further, in Eq.~(\ref{eqn:Wetterich}) we have introduced
\begin{align}
\left(\Gamma_k^{(2)}\right)_{ij}(p,q)=\frac{\delta}{\delta\Phi(-p)}\frac{\delta}{\delta\Phi(q)}\Gamma_k\,.
\end{align}
The FRG can be applied in arbitrary (fractional) dimension. In a variety of cases already low-order truncations have been shown to yield reasonable results, in particular in terms of critical exponents. For reviews on this rapidly evolving method, see, e.g., Refs.~\cite{Berges:2000ew, Polonyi:2001se,Pawlowski:2005xe,Gies:2006wv,Delamotte:2007pf,Rosten:2010vm,Braun:2011pp,metzner2011}.

\subsection{Truncation}

\begin{figure}[!t]
\includegraphics[width=\columnwidth]{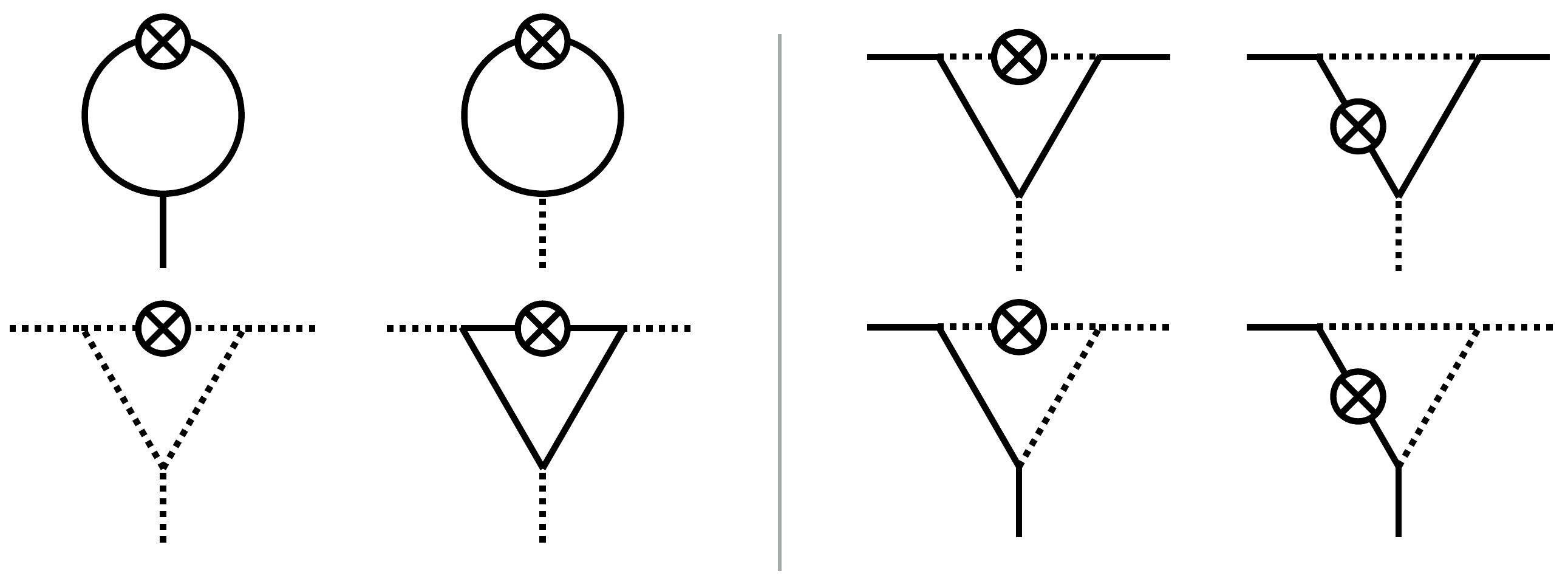}
\caption{\label{alldiags} One-loop diagrams that encode the flow of the couplings within the FRG. Left panel: $\lambda_1$ (upper line) and $\lambda$ (second line). Right panel: $g$ (both lines). We denote $z$  by a dotted line and $\phi$  by a solid line. The loop propagators are regularized nonperturbative propagators, and the regulator insertion $\partial_t R_k$ is denoted by a crossed circle. All diagrams are
evaluated at constant external fields, i.e., vanishing external momenta.}
\end{figure}

The scale-dependent action
contains all possible operators that are compatible with the symmetries. 
The application of the Wetterich equation to a model with an interacting fixed point relies on truncating this space of couplings to a (tractable) subspace. Increasing the order of the truncation then provides a way of testing the reliability and quantitative precision of the results.
In this work, we will consider the following ansatz for the effective action at scale $k$:
\begin{align}\label{eq:truncation}
	\Gamma_k=\int d^dx\Big[\frac{Z_{\phi}}{2}(\partial_\mu \bar \phi)^2+\frac{Z_{z}}{2}(\partial_\mu \bar z)^2+U(\bar\rho,\bar z)\Big]\,,
\end{align}
where we have introduced the field invariant $\bar\rho=\frac{{\bar \phi}^2}{2}$. The wave-function renormalizations $Z_\phi$, $Z_z$ and the effective potential $U$ are scale-dependent quantities.
In the simplest case, one may consider an effective potential of the form
\be
U(\bar{\rho}, \bar z) = \bar{\lambda}_1 \, {\bar z} + \frac{\bar{m}_z^2}{2} {\bar z}^2 + \frac{\bar{\lambda}}{6} {\bar z}^3+ \bar{m}_{\phi}^2\bar{\rho} + \bar{g}\, {\bar z} \, \bar{\rho}.\label{eq:simplesttrunc}
\ee
This ansatz includes the original theory of Eq.~\eqref{eq:lagrangian01} in the limit $Z_z \to 1$, $Z_\phi \to 1$, with the mass parameters sent to $\bar m_z^2\to~0$, $\bar m_\phi^2\to 0$, as well as $\bar \lambda_1\to~0$. 
However, once interaction effects are included, nontrivial momentum dependences can be generated by the RG, a subset of which can be parameterized by the wave-function renormalizations $Z_z$ and $Z_\phi$.
To search for fixed-point solutions, we will work in dimensionless variables:
\begin{gather}
 z = \frac{k^{-(d-2)/2}}{Z_z^{1/2}}\bar z,\quad  \phi = \frac{k^{-(d-2)/2}}{Z_{\phi}^{1/2}}\bar \phi,\\
u(\rho, z) = k^{-d} U(\bar \rho(\rho),\bar z(z)),\quad g=\frac{k^{-3+d/2}}{Z_z\, Z_{\phi}^{1/2}}\, \bar{g}\,.
\end{gather}
The FRG flow equation for the dimensionless renormalized version of the effective potential $u=u(\rho,z)$ reads
\begin{align}
	\partial_t u=&-d u +(d-2+\eta_\phi)\rho\, u^{(1,0)}+\frac{d-2+\eta_z}{2}z\, u^{(0,1)}\nonumber\\
	&+I_{R,\phi}^d(\omega_z,\omega_\phi,\omega_{\phi z})+(N-1)I_{G,\phi}^d(u^{(1,0)})\nonumber\\
	&+I_{R,z}^d(\omega_\phi,\omega_z,\omega_{\phi z})\,,\label{eq:potflow}
\end{align}
where we have introduced dimensionless mass terms in the arguments of the threshold functions $I_{i,j}^d$
\begin{align}
	\omega_\phi&=u^{(1,0)}+2\rho\, u^{(2,0)}\,,\\
	\omega_z&=u^{(0,2)}\,,\\
	\omega_{\phi z}&=\sqrt{2\rho}\, u^{(1,1)}\,.
\end{align}
The threshold functions for the optimized
regulator \cite{litim01} read
\begin{align}
	I_{R,i}^d(x,y,w)&=\frac{4 v_d}{d}\left(1-\frac{\eta_i}{d+2}\right)\frac{1+x}{(1+x)(1+y)-w^2}\,,\\
	I_{G,i}^d(x)&=\frac{4 v_d}{d}\left(1-\frac{\eta_i}{d+2}\right)\frac{1}{1+x}\,,\
\end{align}
where $v_d^{-1}=2^{d+1}\pi^{d/2}\Gamma(d/2)$ and $i=\phi,z$.
For the sharp regulator the threshold functions are%
\footnote{Different definitions for the threeshold functions within the sharp-cutoff scheme are possible, depending on how the sharp-cutoff limit is taken~\cite{Braun:2014wja, Reuter:2001ag}. However, they only differ by a constant and thus do not alter resulting global properties of the potential.}
\begin{align}
	I_{R,i,\text{sh}}^d(x,y,w)&=-v_d\ln\left[(1+x)(1+y)-w^2\right]\,, \displaybreak[0] \\
	I_{G,i,\text{sh}}^d(x)&=-v_d\ln\left[(1+x\right)^2]\,.\
\end{align}
This provides all ingredients required to extract the beta functions for the couplings in Eq.~\eqref{eq:simplesttrunc} from the flow equation \eqref{eq:potflow}. Diagrammatically, these can be encoded in simple one-loop diagrams according to the one-loop structure of the Wetterich equation, cf.~Fig.~\ref{alldiags}.

\begin{figure}[!t]
\includegraphics[width=0.7\columnwidth]{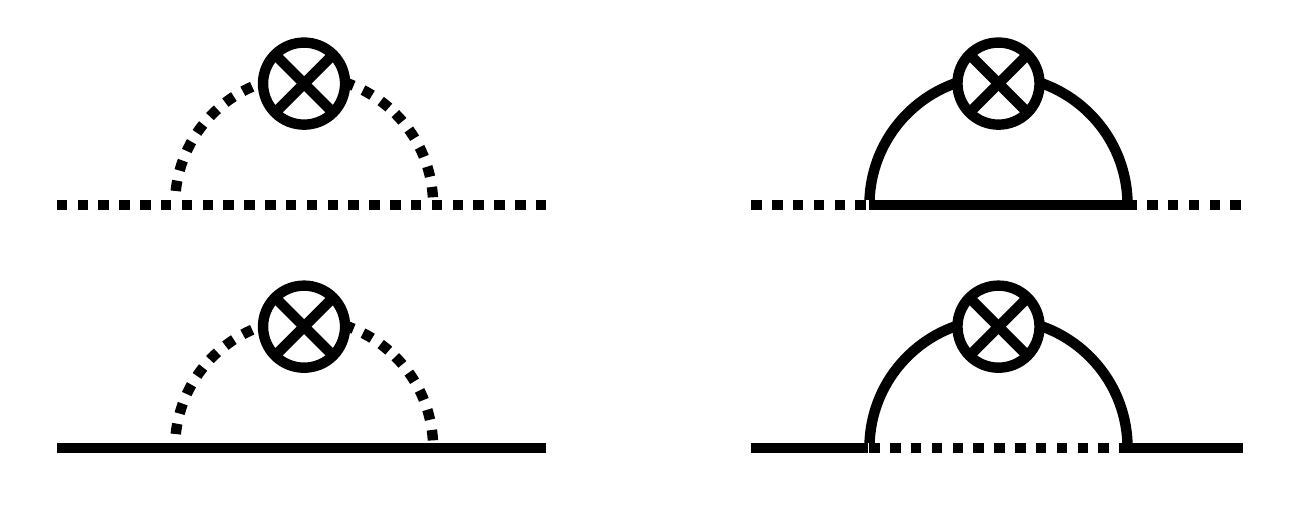}
\caption{\label{etadiags} One-loop diagrams that encode the flow of the mass and the wave-function renormalization for the scalar $z$ (dotted line) and the scalar $\phi$ (solid line), respectively. The loop propagators are regularized nonperturbative propagators, and the regulator insertion $\partial_t R_k$ is denoted by a crossed circle. The anomalous dimensions $\eta_z,\, \eta_{\phi}$ are obtained by expanding to second order in the external momenta, while the beta functions for the masses are encoded in the diagrams at vanishing external momenta.}
\end{figure}

For the anomalous dimensions, encoded in the diagrams in Fig.~\ref{etadiags}, we find for both the flat and the sharp cutoffs
\begin{align}
\eta_z=&\frac{4 v_d}{d}\left( \frac{\lambda^2}{(1+m_z^2)^4}+ N\frac{g^2}{(1+m_{\phi}^2)^4}\right)\,,\label{eq:etaz} \\
\eta_{\phi}=&\frac{8 v_d}{d} \frac{g^2}{(1+m_z^2)^2 (1+m_{\phi}^2)^2}\,.\label{eq:etaphi}
\end{align}
Our conventions for the sharp-cutoff scheme follow those of \cite{Janssen:2012pq}. For $d\rightarrow 6$ and with $m_z=0=m_{\phi}$ in the threshold functions this reduces to the anomalous dimensions given in \cite{Fei:2014yja}, as it should (apart from a factor of 2 which is due to different conventions used in \cite{Fei:2014yja}).

For the effective potential we will employ two different ans\"atze. One suggestion is to treat the Hubbard-Stratonovich transformation similar to the one in many fermionic models and to simply consider an expansion of the dimensionless renormalized effective potential in the $z$ field, i.e.,
\begin{align} \label{eq:usimp}
	u(\rho, z)= ( m_\phi^2 + g  z ) \rho + v( z)\,,
\end{align}
with a function $v \equiv v( z)$ that, for instance, can be expanded in a local potential approximation (LPA). Taking into account the wave function renormalizations or anomalous dimensions, we will refer to such a truncation scheme as LPA${}^\prime$.
This truncation recovers the perturbative fixed-point solution from Ref.~\cite{Fei:2014yja} near $d \nearrow 6$, see the following section.
In practice, we expand  $v( z)$ in a Taylor series to finite order, reading
\begin{align}
	v( z)=\sum_{i=1}^{i_\text{max}}\frac{\lambda_i}{i!}z^i\,,
\end{align}
where $\lambda_2=m_z^2$, $\lambda_3=\lambda$ and $N_\text{max}$ defines the order of the LPA, i.e., LPA$i_\text{max}$. In the limit of $N \to \infty$ we will also study the global properties of $v(z)$ without expanding in the $z$ field, see Sec.~\ref{sec:stability-large-N}.

In Sec.~\ref{sec:twofield} we will  go beyond the ansatz \eqref{eq:usimp} by allowing for higher-order self-interactions of the $\phi$-field, and will refer to this as a ``two-field expansion''.

\section{Extending the fixed point from $d=6-\epsilon$ to $d=5$}
\label{sec:eps-expansion}

We discuss the connection between the FRG and the
$\epsilon$-expansion scheme to show that both approaches coincide in the perturbative limit. This allows us to relate our results to the fixed-point solutions identified in Ref.~\cite{Fei:2014yja}. We will then study these results in detail within the FRG approach by evaluating the fixed-point properties directly in $d=5$ including nonperturbative effects, such as threshold terms in the loop integrals.
%
\begin{figure}[b!]
\includegraphics[width=0.49\columnwidth]{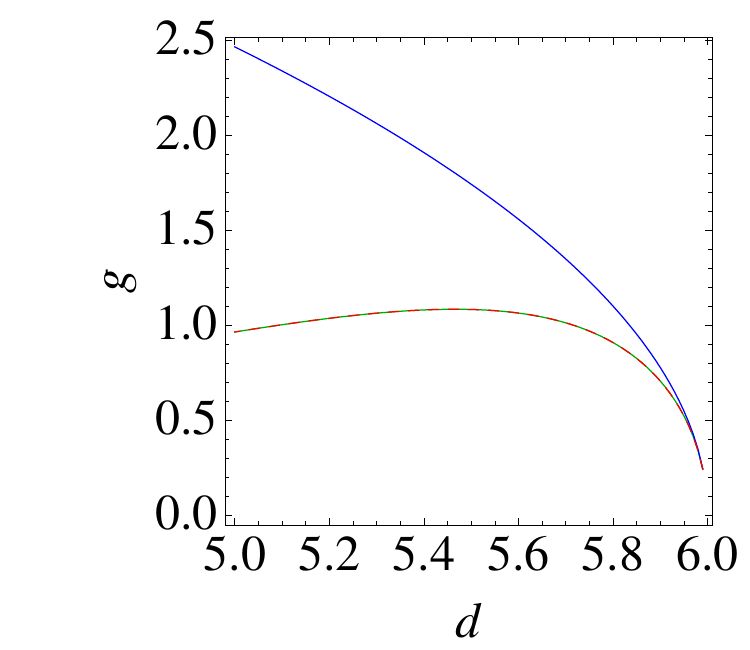}
\includegraphics[width=0.49\columnwidth]{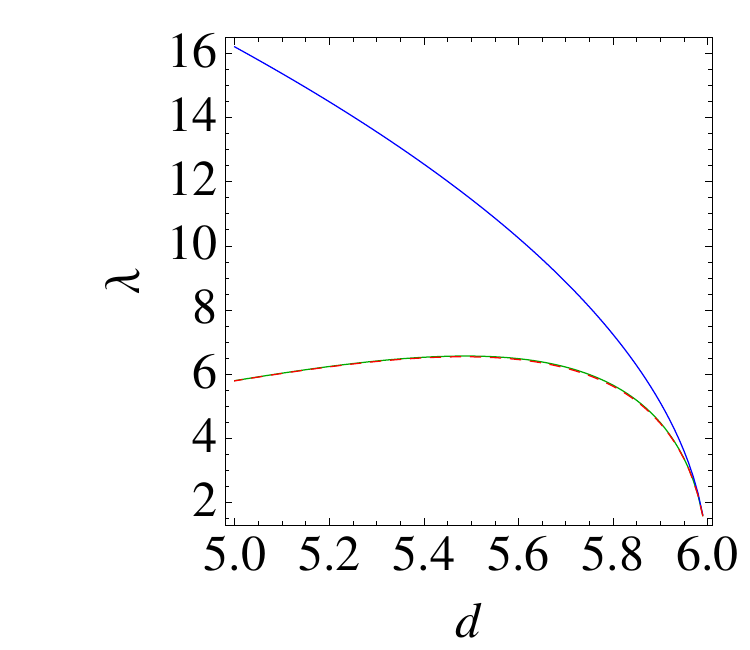}
\includegraphics[width=0.49\columnwidth]{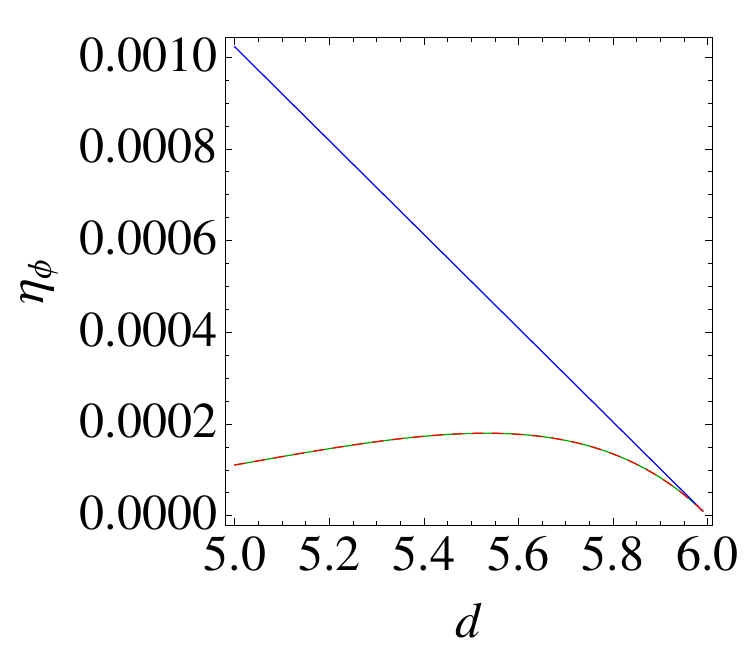}
\includegraphics[width=0.49\columnwidth]{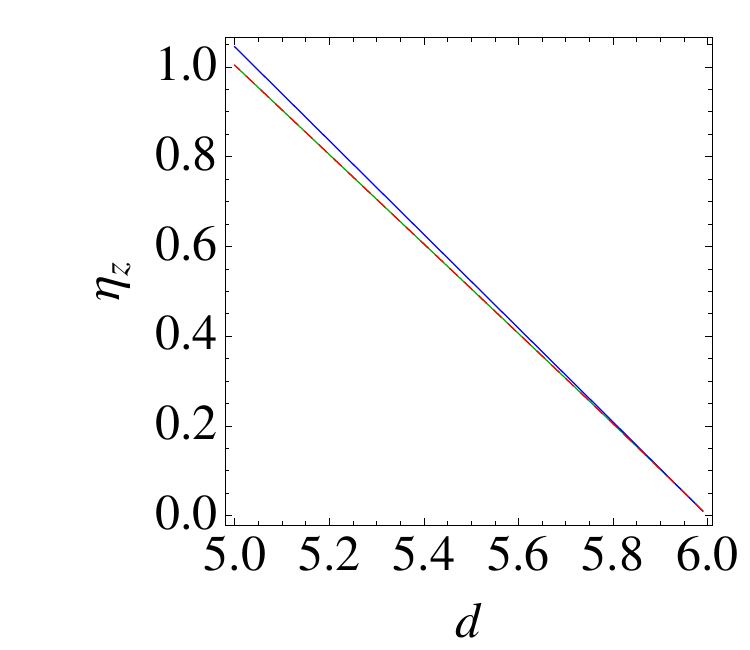}
\caption{Comparison of fixed-point values and anomalous dimensions for $N=2000$ within the $\epsilon$ expansion (blue solid lines) and the FRG in LPA3${}^\prime$ (red dashed) and LPA6${}^\prime$ (green). The different LPA${}^\prime$ results are hardly distinguishable.}
\label{fig:EPSvsFRG}
\end{figure}
%
In the first order of the $\epsilon$ expansion the fixed-point solutions by Fei \emph{et al.}~\cite{Fei:2014yja} appear at large $N>1038$. 
To be explicit, we compare the FRG and the $\epsilon$-expansion results for $N=2000$; however, the agreement we find does not depend on the specific choice of $N$. 
For $d\rightarrow 6$ and to leading order in $6-d$ the results from the FRG and the $\epsilon$ expansion agree perfectly, as expected, cf.~Fig.~\ref{fig:EPSvsFRG}. 
Here, we have worked with the optimized regulator function as this is expected, at least for simple scalar models, to yield the best estimates for the critical exponents \cite{litim,litim01,Pawlowski:2005xe,Pawlowski:2015mlf}. 
For $d \rightarrow 5$ the differences between the two approaches become sizable, which can be attributed to threshold corrections that contribute to the FRG $\beta$ functions.
For better comparison, we also list the values for the couplings and the anomalous dimensions in $d=5.9$ and $d=5.0$ for the FRG and the $\epsilon$ expansion in Tab.~\ref{tab:comp01}. 
Within the FRG, we also show the values from two different truncation schemes (LPA3${}^\prime$ and LPA6${}^\prime$) to display the formidable convergence of the polynomial expansion in the $z$ field. 
For clarity, we display only the stable one of the two fixed points that appear in the $\epsilon$ expansion for $N>1038$, i.e. the one with the lowest number of RG relevant directions.
The other, unstable fixed point which features an additional relevant direction is also found in our FRG approach and coincides with the unstable fixed point from $\epsilon$ expansion in the limit $d\rightarrow 6$. 

\begin{table}[t!]
\caption{\label{tab:comp01} Fixed point values and anomalous dimensions for $N=2000$ in $d=5.9$ and $d=5$.}
\begin{tabular*}{\linewidth}{@{\extracolsep{\fill} } c c c c c c}
\hline\hline
approximation & $d$ & $g$ & $\lambda$ &$\eta_\phi$  & $\eta_z$\\
\hline
 $\epsilon^1$ exp & 5.9 & 0.7801 & 5.1215 & 0.00010223 & 0.10444 \\
FRG LPA3${}^\prime$& 5.9 & 0.7089 & 4.4708 & 0.00008289 & 0.10327 \\
FRG LPA6${}^\prime$& 5.9 & 0.7089 & 4.4956 & 0.00008288 & 0.10329 \\
\hline
$\epsilon^1$ exp & 5 & 2.4670 & 16.1957 & 0.00102233 & 1.04436 \\
FRG LPA3${}^\prime$& 5 & 0.9664 & 5.7938 & 0.00011043 & 1.00338 \\
FRG LPA6${}^\prime$ & 5 & 0.9664 & 5.7993 & 0.00011040 & 1.00338\\
\hline\hline
\end{tabular*}
\end{table}

\section{Critical $N$ from the FRG}
\label{sec:critical-N}

In the $\epsilon$ expansion the stable fixed point was found to exist only above a certain critical $N_c$, with $N_c$ (beyond leading order) depending on the dimension $d$. To leading order the result is independent of $d$ and one finds $N_c\approx 1038$.
As has been emphasized, e.g., in \cite{Fei:2014xta}, it is highly desirable to use a nonperturbative method to obtain another estimate of $N_c$ beyond the $\epsilon$ expansion. 
We now compute $N_c (d)$ within the truncation Eq.~\eqref{eq:truncation} with Eq.~\eqref{eq:usimp} of the functional renormalization group.

\begin{table}[b!]
\caption{\label{tab:smallN} The fixed-point values and anomalous dimensions for $N=1$, $N=2$ and $N=10$ in $d=5$ exhibit convergence at increasing orders of the local potential approximation (LPA${}^\prime$).}
\label{tab:I+H}
\begin{tabular*}{\linewidth}{@{\extracolsep{\fill} } c c c c c c}
\hline\hline
approximation & $N$ & $g$ & $\lambda$ &$\eta_\phi$  & $\eta_z$\\
\hline
FRG LPA3${}^\prime$& 1 & 70.686 & 346.221 & 0.00899 & 1.8108\\
FRG LPA4${}^\prime$& 1 & 70.632 & 347.484 & 0.00897 & 1.8096\\
FRG LPA5${}^\prime$& 1 & 70.622 & 350.745 & 0.00895 & 1.8097 \\
FRG LPA6${}^\prime$& 1 & 70.622 & 350.777 & 0.00895 & 1.8097 \\
\hline
FRG LPA3${}^\prime$& 2 & 46.411 & 231.184 & 0.01184 & 1.6921 \\
FRG LPA4${}^\prime$& 2 & 46.342 & 233.323 & 0.01175 & 1.6898 \\
FRG LPA5${}^\prime$& 2 & 46.336 & 236.466 & 0.01171 & 1.6903 \\
FRG LPA6${}^\prime$& 2 & 46.334 & 236.541 & 0.01170 & 1.6902 \\
\hline
FRG LPA3${}^\prime$ & 10 & 16.802 & 90.398 & 0.01013 & 1.3474 \\
FRG LPA4${}^\prime$ & 10 & 16.783 & 93.060 & 0.00993 & 1.3464 \\
FRG LPA5${}^\prime$ & 10 & 16.787 & 94.725 & 0.00988 & 1.3477 \\
FRG LPA6${}^\prime$ & 10 & 16.786 & 94.870 & 0.00987 & 1.3477 \\
\hline\hline
\end{tabular*}
\end{table}

\begin{figure}[t!]
\includegraphics[width=0.49\columnwidth]{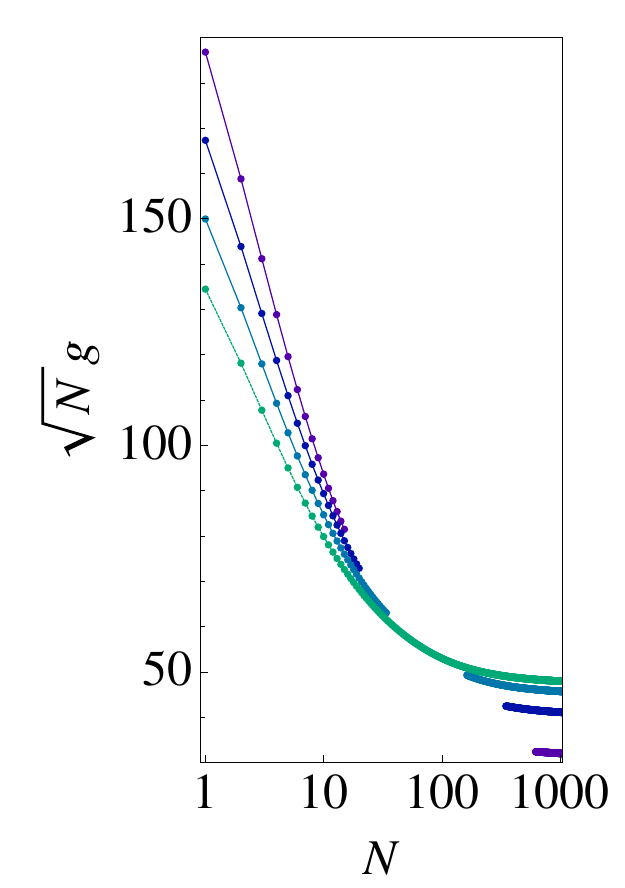}
\includegraphics[width=0.49\columnwidth]{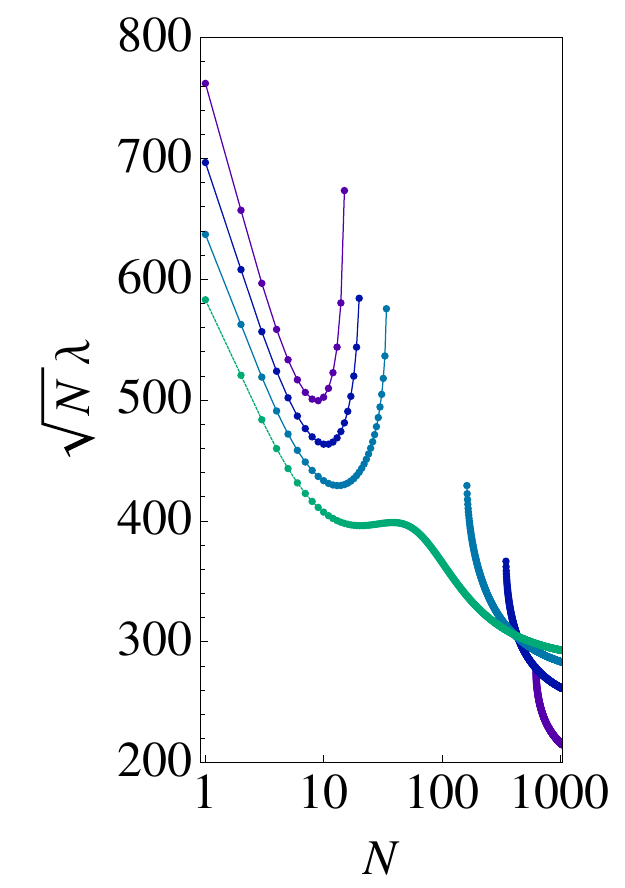}
\includegraphics[width=0.49\columnwidth]{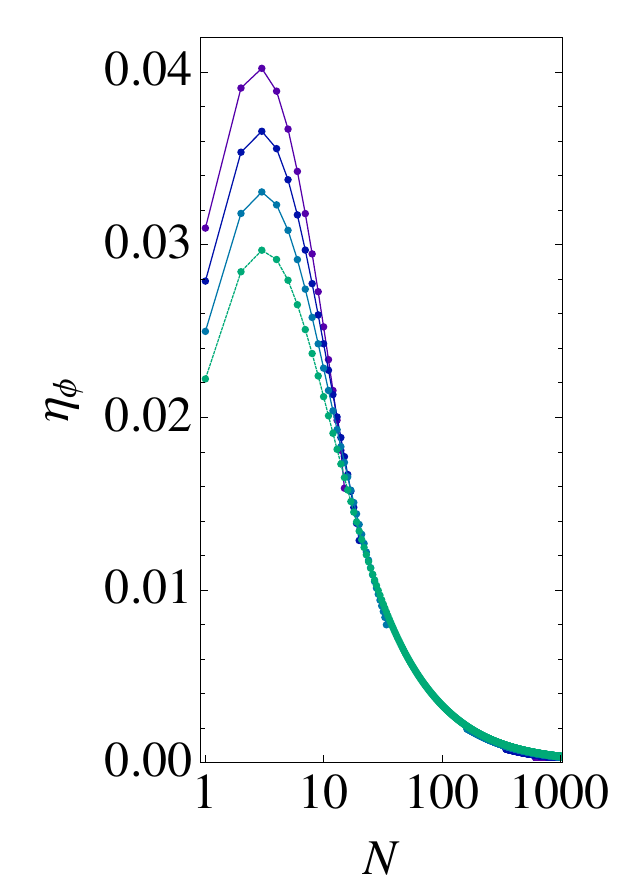}
\includegraphics[width=0.49\columnwidth]{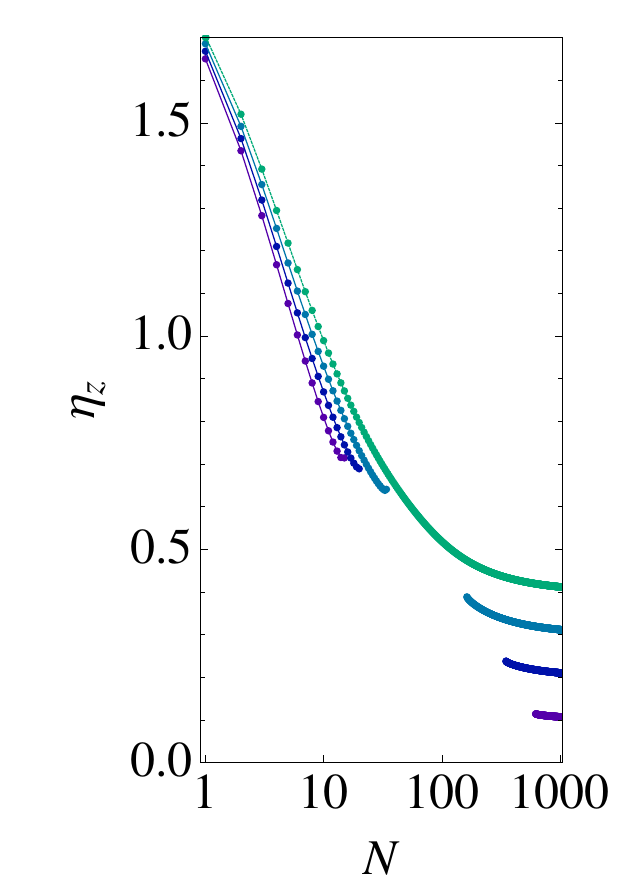}
\caption{\label{FP_N_d5a} Fixed-point values and anomalous dimensions in LPA3${^\prime}$ for $d \gtrsim d_c$ ($d=5.9, 5.8, 5.7,5.6$ from purple to green). For these choices of $d$, the large-$N$ fixed point as discovered from the $\epsilon$ expansion~\cite{Fei:2014yja} can be continued up to some critical $N_c(d)$ with $N_c(d \to 6) \approx 1038$, $N_c(5.9)\approx 623$, $N_c(5.8)\approx 349$, and $N_c(5.7)\approx 163$. However, we find that the fixed point reappears on the real axis at small values of $N$ and large values of the couplings $\sim \mathcal O(1)$, even in the limit $d\to 6$ (``small-$N$ island'').}
\end{figure}

In particular, we are interested in determining whether a minimal value of $N$, below which the fixed point ceases to be real, exists within the FRG setting. 
As is obvious from Fig.~\ref{fig:EPSvsFRG}, our results are close to those of the leading-order $\epsilon$ expansion for $d \to 6$. 
Accordingly, our numerical evaluation indeed yields a critical value of $N$, at which the stable fixed point merges with the unstable fixed point, and both disappear into the complex plane. 
For $d=5.999$ we get $N_c \approx 1034$, while for $d=5.9$ we already find $N_c \approx 623$.
Towards lower dimensionality, the critical value decreases, and we find $N_c \approx 349$ in $d=5.8$ and  $N_c \approx 163$ in $d=5.7$, cf.~Fig.~\ref{FP_N_d5a}, in qualitative agreement with the $\epsilon$-expansion results.

\begin{figure}[t!]
\includegraphics[width=0.49\columnwidth]{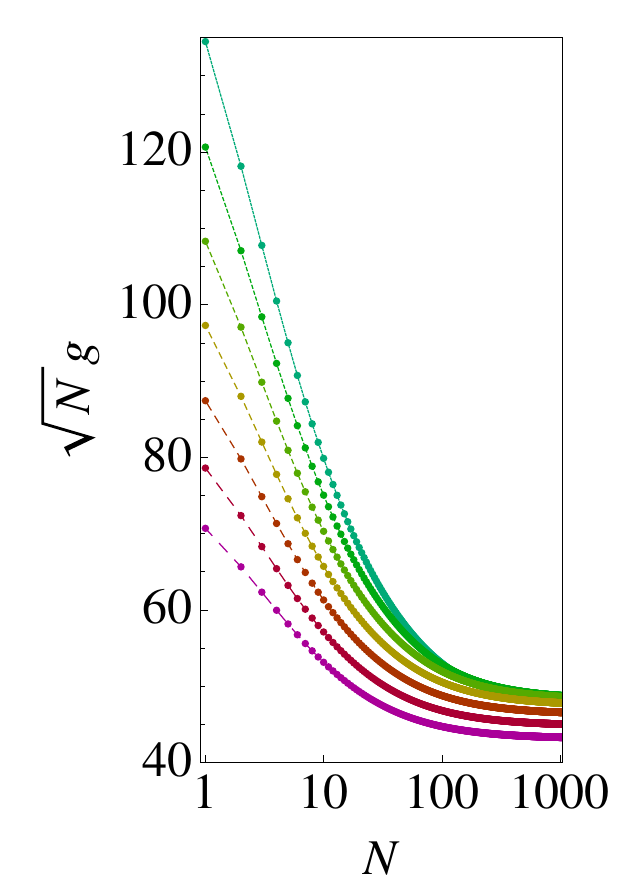}
\includegraphics[width=0.49\columnwidth]{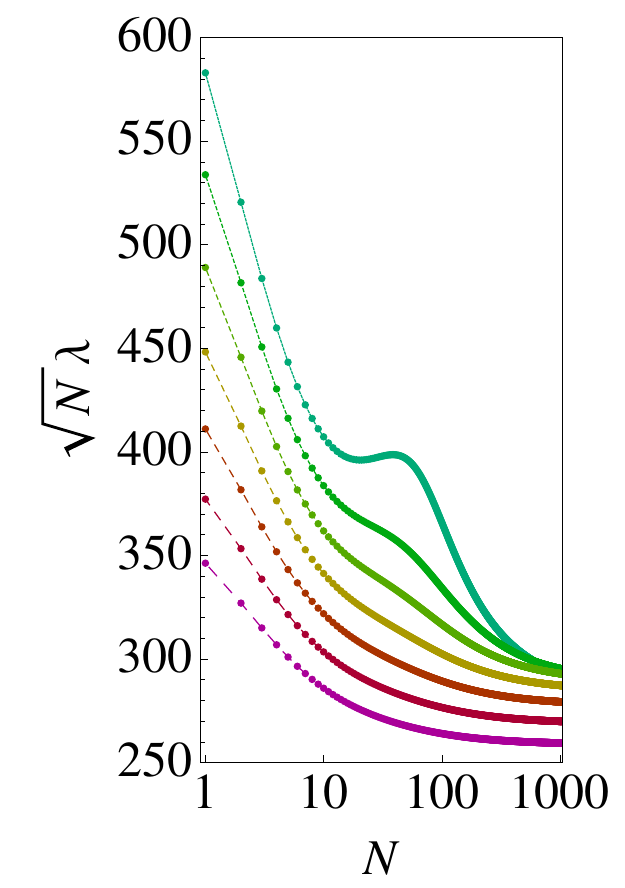}
\includegraphics[width=0.49\columnwidth]{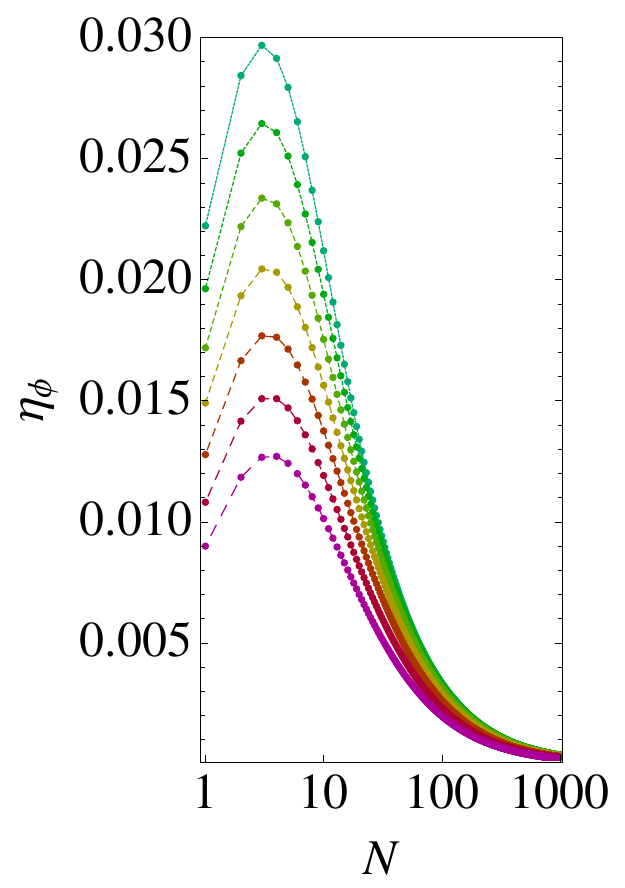}
\includegraphics[width=0.49\columnwidth]{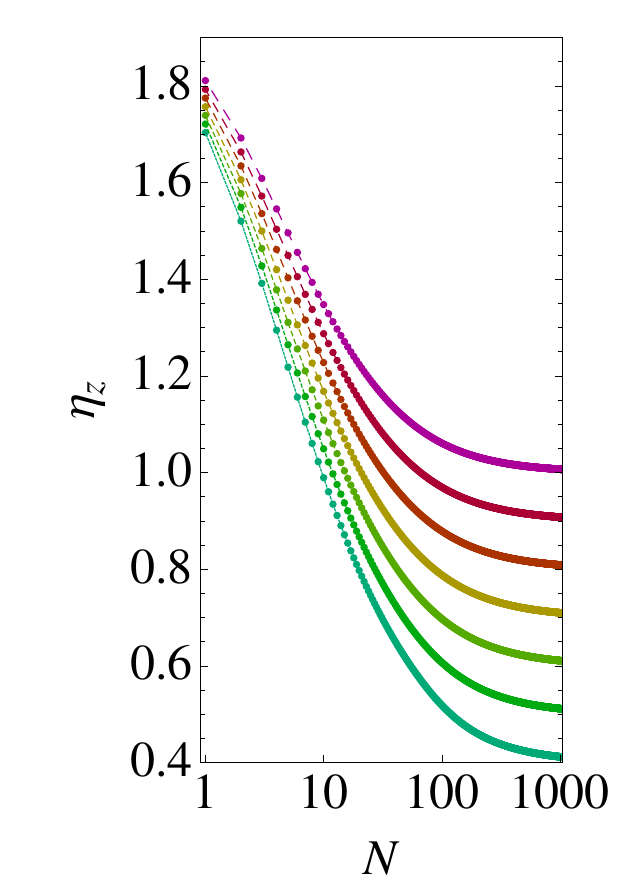}
\caption{\label{FP_N_d5} Fixed-point values and anomalous dimensions for $d<d_c$ ($d=5.6,5.5,5.4,5.3,5.2,5.1,5.0$ from green to red). For these choices of $d$ the large-$N$ FRG fixed-point solution can be continued at real values of the couplings all the way down to $N=1$.}
\end{figure}

\begin{figure}
\includegraphics[width=0.8\columnwidth]{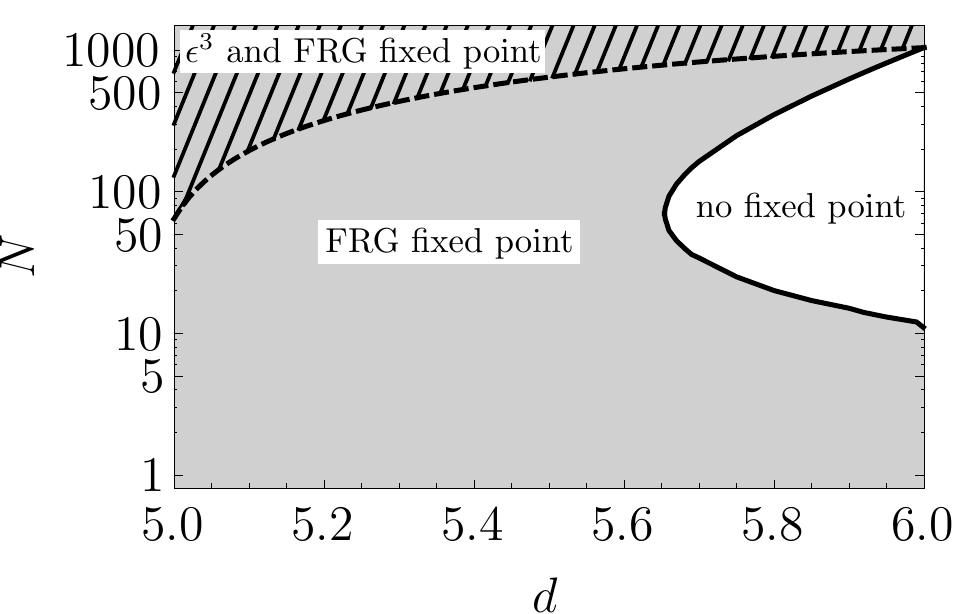}
\caption{\label{Ndexistence} Region of existence of a stable and real fixed within FRG in LPA3${}^\prime$ (gray shaded) and, for comparison, from the third-order $\epsilon$ expansion (hatched)~\cite{Fei:2014xta}. The values for $N_c$ (solid and dashed line, respectively) agree in the vicinity of $d=6$, as they should, but then deviate significantly, due to threshold effects in the FRG flow. Moreover, we find within the FRG that the fixed point reappears on the real axis for small values of $N$ in the limit $d \to 6$.}
\end{figure}

As a difference to the $\epsilon$ expansion, we find that the fixed point reappears on the real axis for lower values of $N$, see Fig.~\ref{FP_N_d5a}. 
Near and below $d=6$, there are two ``islands'' on the $N$ axis on which the fixed point is real. Below some noninteger \emph{critical} dimension $d_c$, we find that
these two islands merge, and the fixed point exists at real values of the couplings for \emph{all} $N\geq 1$, see Fig.~\ref{FP_N_d5}.
%
%
Within our approximation we obtain $d_c \approx 5.65$ numerically.
Our results thus suggest that a critical value of $N$ does not exist in the physical situation of $d=5$.
A summary of these findings is shown in Fig.~\ref{Ndexistence} and the convergence of the results within the LPA${}^\prime$ is exhibited in Tab.~\ref{tab:I+H}.
%

The existence of a second island for low $N$ and $d>d_c$ in which the fixed point reappears on the real axis is due to nonperturbative threshold effects of our FRG equations: 
In fact, we find that the fixed-point couplings in this ``low-$N$ island'' have values of $\mathcal{O}(1)$, even when $d\to 6$. 
This second island is therefore inaccessible by the perturbative expansion, even when $\epsilon$ is small.
The reappearance of the fixed point at low $N$ near $d \to 6$ is in fact responsible for the occurrence of the noninteger critical dimension $d_c$, and hence ultimately for the comparatively large quantitative disagreement between our FRG estimates and the values obtained from extrapolating the perturbative expansion to larger values of $\epsilon$.
While at present we cannot exclude the possiblity that the reappearance of the fixed point at low $N$ is an artifact of our truncation, we 
do not find any sign of breakdown of our approximation at low $N$. A similar ``low-$N$ island'' has, in fact, recently been identified within the $6-\epsilon$ expansion of a related cubic $O(N)$ theory in which the Hubbard-Stratonovich decoupling is made in the \emph{tensorial} channel as opposed to the scalar channel~\cite{Herbut:2015zqa}. 
It is certainly tempting to identify these small-$N$ fixed points with our low-$N$ island. However, 
understanding the relation between the fixed points in the tensorial and scalar cubic models 
requires
to overcome the ambiguity in the different 
Hubbard-Stratonovich decouplings of the $\phi^4$ interaction (so-called Fierz or mean-field ambiguity)~\cite{Jaeckel:2002rm}. Within the FRG,  
this would be possible by means of dynamical bosonization~\cite{Gies:2001nw, Pawlowski:2005xe, Floerchinger:2009uf}. This is left for future work.

The existence of a noninteger critical dimension $4<d_c<6$ below which the fixed point 
exists for all $N\geq 1$ is in fact requisite to the conjecture that the fixed point of the cubic scalar theory in $d=6-\epsilon$ represents the analytic continuation of the $O(N)$ Wilson-Fisher fixed point towards $d>4$~\cite{Fei:2014yja}, as well as to the case where the $O(N)$ universality class is embedded in that of our model, see Sec.~\ref{sec:universality-classes}.
In $d = 4+\epsilon$ the Wilson-Fisher fixed point can be accessed by the $\epsilon$ expansion of a pure $(-\phi^4)$ theory and hence is guaranteed to exist (though at negative quartic coupling) for all values of $N\geq 1$ as long as $\epsilon$ is small.
If the fixed point of our cubic model defines the cubic $O(N)$ universality class in $d>4$ one therefore expects already on general grounds 
that there exists a critical dimension $d_c$ below which no merging and annihilation of fixed points occurs, with $4<d_c<6$.
The main question is only whether $d_c$ is above or below the physical dimension.
Within our approximation we find $d_c \approx 5.65$ and thus \emph{above} $d=5$.
It should in principle be possible 
to test this scenario with standard Monte Carlo simulations.
We conjecture that such simulations should display universal critical behavior governed by our stable fixed point for any value of $N$ in $d=5$.
We should emphasize, however, that it is unclear at present, whether a lattice formulation (with positive measure) exists, which allows to demonstrate the existence of our fixed point and assess its properties. We further discuss this problem in the concluding section.

Overall, the situation appears to be similar to the issue of the fixed-point structure of the three-dimensional Abelian Higgs model
\cite{coleman1973, halperin1974,dasgupta1981, kolnberger1990,Bergerhoff:1995zm,bergerhoff1996,herbut1996,herbut1997,Freire:2000sx}:
Within the $\epsilon$ expansion in $d=4-\epsilon$ dimensions, one finds a critical number $N^{\rm (AH)}_c$ of complex scalars below which a real and stable fixed point ceases to exist, with $N_{c}^{\rm (AH)} \approx 183$ to leading order in the $\epsilon$ expansion~\cite{coleman1973, halperin1974}. However, already the next-to-leading order term drastically reduces the estimate when extrapolated to $\epsilon = 1$~\cite{kolnberger1990, herbut1997}. It has therefore been speculated that the standard $\epsilon$ expansion may be inapplicable to the physically relevant situation in $d=3$, forcing one to go back to alternative
approximations~\cite{herbut1997}. 
The perturbative RG analysis in fixed $d=3$~\cite{herbut1996}, in agreement with earlier functional RG studies~\cite{bergerhoff1996}, predicted that a stable and real fixed point may in fact continue to exist all the way down to $N^{\rm (AH)}=1$. This scenario is also consistent with the data from lattice Monte Carlo simulations~\cite{dasgupta1981}. It is therefore nowadays believed that in fact $N_{c}^{\rm (AH)}<1$ in $d=3$, the lowest-order $\epsilon$-expansion result notwithstanding~\cite{herbut2007}.

\section{Comparison of universality classes}
\label{sec:universality-classes}

A universality class is characterized by universal scaling exponents. These include the anomalous dimensions of the fields, as well as the exponents that characterize the linearized RG flow of the essential couplings around a fixed point. Within a RG scheme that includes beta functions for the masses, all critical exponents can be obtained by diagonalizing the stability matrix in the vicinity of the fixed point. Specifically, denoting the set of all dimensionless couplings, including the masses, by $\{ g_i\}$, with fixed-point values $\{g_{i\,\ast}\}$, the stability matrix $\mathcal{M}$ is given by 
\be
\mathcal{M}_{mn} = \frac{\partial \beta_{g_m}}{\partial g_n} \Big|_{g_i= g_{i\, \ast}}.
\ee
The critical exponents are defined as the eigenvalues of this matrix, multiplied by a negative sign, 
\be
\theta_I = - {\rm eig}\, \mathcal{M}_{mn}.
\ee
In the vicinity of the Gaussian fixed point, the critical exponents agree exactly with the canonical dimensions of the couplings, whereas residual interactions result in additional contributions near an interacting fixed point. While beta functions of dimensionful couplings are non-universal, i.e., scheme-dependent, the critical exponents are scheme-independent. Thus the set of critical exponents is determined uniquely by the symmetries, dimensionality and degrees of freedom of a system.%
\footnote{Exceptions to this general rule have been discussed in Ref.~\cite{Gehring:2015vja}.}

We will explicitly compare the critical exponents for the pure quartic $O(N)$ model (analytically continued to $d>4$) with those of the cubic model specified in Eq.~\eqref{eq:lagrangian01}, in order to explore whether the universality classes agree. If they do not, then the pure $O(N)$ model and the model that we consider here are actually inequivalant on the quantum level.

\subsection{Analytic continuation of the Wilson-Fisher $O(N)$ fixed point to $d>4$}

In the large-$N$ limit, solutions of the flow equation for the effective potential within the original quartic $O(N)$ formulation are available \cite{Litim:1995ex,Tetradis:1995br,Mati:2016wjn}, and yield \cite{Mati:2016wjn}
\be\label{eq:theta1a}
\nu = \frac{1}{\theta_1}= \frac{1}{d-2},
\ee
in agreement with the simple analytic continuation of the result in $d<4$~\cite{ZinnJustin:1998cp}.
Similarly, the second-largest critical exponent of the large-$N$ fixed point is determined by
\bee\label{eq:theta2a}
\omega &=& -\theta_2 = 4-d,
\eee
again continuing the $d<4$ result from Ref.~\cite{ZinnJustin:1998cp} towards $d>4$.
The Wilson-Fisher fixed point in $d>4$ thus has \emph{two} relevant directions ($\theta_2 > 0$), in contrast to the case of $d<4$.
The additional relevant direction is associated with the quartic coupling, which has canonical scaling dimension $4-d$, and is therefore perturbatively irrelevant at the Gaussian fixed point. 
Thus the Gaussian fixed point is IR attractive in the quartic coupling. 
Accordingly a non-Gaussian fixed point will be UV attractive, i.e., IR relevant in that coupling. 
The other relevant direction is associated with the mass operator, and has canonical dimension $2$ independent of $d$.
  
\subsection{Universality class of the cubic $O(N)$ model}
 
For the model defined in Eq.~\eqref{eq:lagrangian01} with the Hubbard-Stratonovich field $z$ we find
 the anomalous dimensions in the limit $N\rightarrow\infty$:
\begin{align}\label{eq:anomN}
N\rightarrow \infty \text{:}\qquad\eta_\phi\rightarrow 0, \qquad \eta_z\rightarrow 6-d,
\end{align}
cf.~Figs.~\ref{FP_N_d5a}, \ref{FP_N_d5} and Sec.~\ref{sec:stability-large-N}. The scaling dimensions for the fields $\phi$ and $z$ are thus
\bee
\Delta_{\phi} &=& \frac{d}{2}-1 + \frac{\eta_{\phi}}{2} \overset{N \rightarrow \infty}{\longrightarrow} \frac{d}{2}-1,\\
\Delta_z &=&  \frac{d}{2}-1 + \frac{\eta_{z}}{2} \overset{N \rightarrow \infty}{\longrightarrow} 2,
\eee
in agreement with Ref.~\cite{Fei:2014yja}.

The beta functions for the two mass parameters in the simplest truncation specified by Eq.~\eqref{eq:simplesttrunc} in LPA2${}^\prime$, i.e., $\lambda\equiv 0$, read
\begin{align}
 \beta_{m_z^2} & = (-2+ \eta_z) m_z^2+ g^2\frac{8 v_d}{d} N \frac{1- \frac{\eta_{\phi}}{2+d}}{(1+m_{\phi}^2)^3}, \\
 \beta_{m_\phi^2} & = (-2+\eta_{\phi})m_{\phi}^2 + g^2\frac{8 v_d}{d} \frac{1}{(1+m_{\phi}^2)(1+m_z^2)} \nonumber\\
 & \quad \quad \quad\quad\quad\quad \times \left(\frac{1- \frac{\eta_z}{2+d}}{1+ m_z^2} + \frac{1- \frac{\eta_{\phi}}{2+d}}{1+m_{\phi}^2} \right)\,.
\end{align}
Further, the beta function for $g$ is
\begin{align}
 \beta_g & = \frac{1}{2}\left(d-6+\eta_z+2\eta_{\phi}\right)g + g^3\frac{8 v_d}{d} \frac{1}{(1+m_{\phi}^2)^2 (1+ m_z^2)} \nonumber \\
 & \quad\quad \quad \quad \quad \quad \quad\quad\quad\quad \times \left( - \frac{1- \frac{\eta_z}{d+2}}{1+ m_z^2} - 2\frac{1- \frac{\eta_{\phi}}{d+2}}{1+m_{\phi}^2}\right).\label{eq:betag}
\end{align}
To obtain all critical exponents, it is crucial to consider also the renormalization of the operator linear in $z$, i.e., the $\beta$ function for $\lambda_1$:
\bee
\beta_{\lambda_1}&=&  \left(1-\frac{d}{2}+ \frac{\eta_z}{2} \right)\lambda_1 - g\frac{4 v_d}{d}N\frac{1 - \frac{\eta_{\phi}}{d+2}}{(1+m_{\phi^2})^2}.
\eee
Including the expressions for the anomalous dimensions, Eq.~\eqref{eq:etaz} and \eqref{eq:etaphi}, we find the interacting fixed point, which, e.g., in the case of $d=5$, $N=2000$ lies at
\begin{align}
m_{\phi\, \ast}^2 & =2.14\cdot 10^{-4}, & m_{z\, \ast}^2 & = 2.006, \nonumber \\
g_{\ast} & = 0.965, & \lambda_{1\, \ast} & = -0.346.   
\end{align}
Inserting these fixed-point values in the expression for the stability matrix yields the critical exponents $\theta_1 =2.999$, $\theta_2 = 0.998$ and $\theta_3=2.002$ and $\theta_4=-1.001$.
The result for $\theta_2$ and $\theta_3$ can be elucidated by the following argument: As the interacting fixed point that we consider emanates from the Gaussian fixed point in $d=6$, we can assume that the critical exponents correspond to scaling dimensions of the couplings $m_{\phi}^2$ and $m_z^2$ at the fixed point. 
For these, the following equations hold:
\begin{align}
 [m_z^2] =& 2- \eta_z = d-4, \label{eq:theta2b}\\
 [m_{\phi}^2]=& 2 - \eta_{\phi} =2 \label{eq:theta3b},
\end{align}
where we have inserted $\eta_{\phi} =0$ and $\eta_z= 6-d$, which holds in the large-$N$ limit. This argument also motivates this particular value for $\eta_z$: The ``composite" field $z$ must scale in that particular way, in order to reproduce scaling exponents from the pure quartic $O(N)$ model.
Our explicit results for $\theta_{2,3}$ for $d=5$ and $N=2000$ approach the values in Eq.~\eqref{eq:theta2b} and \eqref{eq:theta3b} very closely. This confirms that our reasoning does indeed explain the observed critical exponents: Essentially, the scaling at the interacting fixed point is the canonical scaling, shifted by the anomalous dimension for the $z$ field. The largest critical exponent $\theta_1$ can be understood as the anomalous scaling of the operator $\lambda_1 z$ present in Eq.~\eqref{eq:simplesttrunc}, which has scaling dimension $(d+2-\eta_z)/2 = d-2$.
The explicit result in $d=5$ already highlights the existence of three relevant directions, i.e., positive critical exponents, in contrast to only two for the pure quartic $O(N)$ model. This is true in any dimension $4<d<6$, cf.~Tab.~\ref{thetatab}, where we also include the self-interaction of the $z$ field (LPA3'). There, we use a notation in which the largest critical exponent, $\theta_1$, is related to $\nu_{\rm FRG}$ by $\theta_1 = 1/\nu_{\rm FRG}$. The second-largest critical exponent is denoted $\theta_3$. The third-largest critical exponent is related to $\omega_{\rm FRG}$ by $\theta_2 = - \omega_{\rm FRG}$.

\begin{figure}[t!]
\includegraphics[width=0.9\columnwidth]{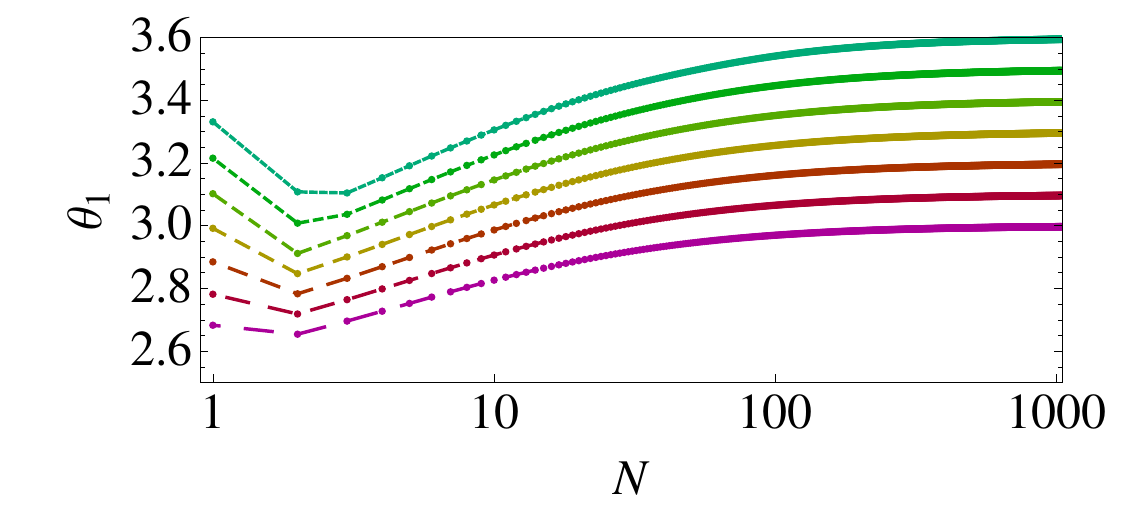}
\includegraphics[width=0.9\columnwidth]{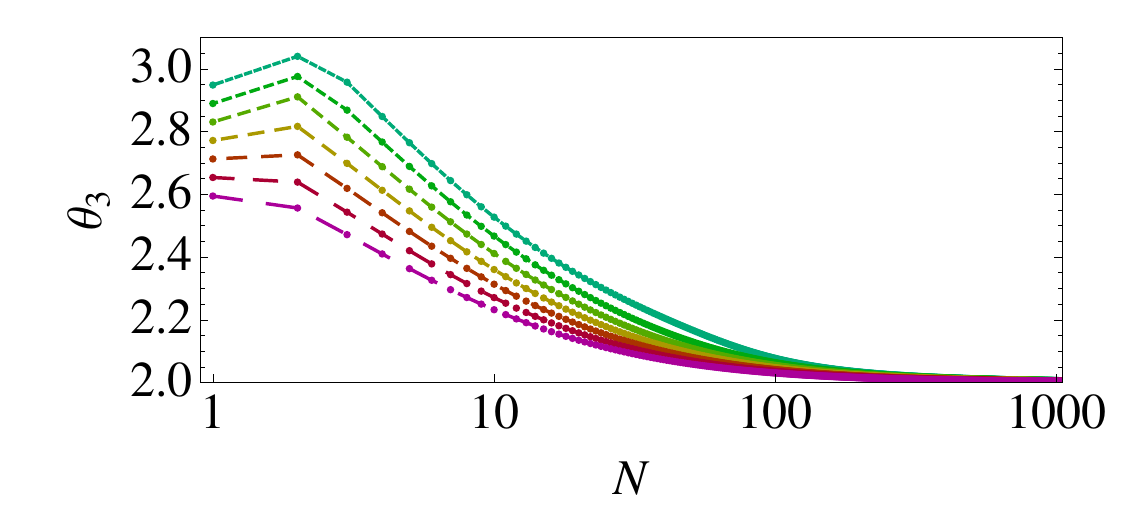}
\caption{\label{FP_N_d52} Critical exponents $\theta_{i}$. From green to red following the order of the colors of the rainbow for $d=5.6,5.5,5.4,5.3,5.2,5.1,5.0$. For large $N$ the exponents $\theta_1 \rightarrow d-2$ and $\theta_2\rightarrow d-4$ as given in Eqs.~\eqref{eq:theta1a} and \eqref{eq:theta2a}.}
\end{figure}

We have plotted the critical exponents $\theta_i, i \in \{1,2,3\}$ for various dimensions $d\leq 5.6$ as a function of $N$ in Fig.~\ref{FP_N_d52}. At large $N$ this confirms the expected behavior, where $\theta_1 \rightarrow d-2$, $\theta_2 \rightarrow d-4$ and $\theta_3\rightarrow 2$.
 
 \begin{table}[!top]
 \caption{\label{thetatab} Critical exponents for $N=2000$ in FRG with LPA3${}^\prime$ and in large-$N$ limit of the analytic continuation of the Wilson-Fisher $O(N)$ fixed point to $d>4$. The departure of the FRG values from the large-$N$ results is exactly of the size $1/2000$, as one would expect. This suggests that our FRG results coincides with the large-$N$ expansion in the limit $N\to \infty$ and the FRG fixed point indeed describes the $O(N)$ universality class above four dimensions.}
 \begin{tabular*}{\linewidth}{@{\extracolsep{\fill} } c c c c c c c}
 \hline\hline
 $d$ & $N$ & $\nu_{\rm O(N)}$ & $\omega_{\rm O(N)}$ & $\nu_{\rm FRG}$ & $\omega_{\rm FRG}$ & $\theta_{3\, FRG}$ \\\hline
 5.9 & 2000 & 0.256 & -1.9 & 0.257 & -1.904 & 2.002\\ 
 5.8 & 2000 & 0.263 & -1.8 & 0.263 & -1.805 & 2.002\\ 
 5.7 & 2000 & 0.270 & -1.7 & 0.270 & -1.705 & 2.003\\ 
 5.6 & 2000 & 0.278 & -1.6 & 0.278 & -1.604 & 2.003\\ 
 5.5 & 2000 & 0.286 & -1.5 & 0.286 & -1.504 & 2.003\\ 
 5.4 & 2000 & 0.294 & -1.4 & 0.294 & -1.403 & 2.003\\ 
 5.3 & 2000 & 0.303 & -1.3 & 0.303 & -1.302 & 2.002\\ 
 5.2 & 2000 & 0.313 & -1.2 & 0.313 & -1.202 & 2.002\\ 
 5.1 & 2000 & 0.326 & -1.1 & 0.323 & -1.101 & 2.002\\ 
 5.0 & 2000 & 0.333 & -1.0 & 0.334 & -1.003 & 2.002\\
 4.9 & 2000 & 0.345 & -0.9 & 0.345 & -0.901 & 2.001\\
 4.8 & 2000 & 0.357 & -0.8 & 0.357 & -0.800 & 2.001\\
 4.7 & 2000 & 0.370 & -0.7 & 0.371 & -0.700 & 2.001\\ 
 4.6 & 2000 & 0.385 & -0.6 & 0.385 & -0.600 & 2.001\\ 
 4.5 & 2000 & 0.400 & -0.5 & 0.400 & -0.500 & 2.001\\
 4.4 & 2000 & 0.417 & -0.4 & 0.417 & -0.400 & 2.001\\ 
 4.3 & 2000 & 0.435 & -0.3 & 0.435 & -0.300 & 2.000\\
 4.2 & 2000 & 0.455 & -0.2 & 0.455 & -0.200 & 2.000\\
 4.1 & 2000 & 0.476 & -0.1 & 0.476 & -0.100 & 2.000\\ 
 \hline\hline
 \end{tabular*}
\end{table}

As discussed in Ref.~\cite{Fei:2014yja}, the operators $z^3$ and $z \phi^2$ will feature couplings with scaling dimension $\epsilon = 6-d$ at large $N$. We are able to confirm this behavior within our approach, for instance at $N=2000$, where $\theta_4 = -0.496, \, \theta_5 =-0.501$ at $d=5.5$ and $\text{Re}[\theta_4]= -1.003, \, \text{Re}[\theta_4] = -1.003$ at $d=5$.

The excellent agreement of the critical exponents $\nu$ and $\omega$ between the quartic $O(N)$ model and the cubic $O(N)$ model shows that the corresponding fixed points define related universality classes. Note that it is crucial to include the linear term $\lambda_1 \, z$ into the flowing action to obtain a critical exponent that agrees with $\nu = \frac{1}{d-2}$ from the pure quartic $O(N)$ model. On the other hand, there is an additional relevant direction in the cubic model, which is related to the mass-parameter of $\phi$, cf.~Eq.~\eqref{eq:theta3b}. The universality class of the quartic $O(N)$ model is thus \emph{embedded} into that of the cubic model: After tuning the third relevant direction, the remaining tuning that is required in order to reach the fixed point in the IR is exactly the same in both models. Towards the UV, there is a third attractive direction of the fixed point, i.e., the critical hypersurface is three-dimensional.
A similar behavior occurs in another class of models which contain multiple scalar fields: Coupling an $O(N)$ vector and an $O(M)$ vector in $d=3$ results in a variety of fixed points, one of which is the so-called isotropic fixed point (IFP), at which the symmetry is enhanced to an $O(N+M)$ symmetry~\cite{herbut2007}. Thus, the critical exponents of the simple $O(N+M)$ model can be recovered from the IFP. On the other hand, it features (at least) one additional relevant direction -- again, the simpler universality class is embedded in the universality class of the model with additional fields \cite{Eichhorn:2013zza}. The same structure can be observed in models with even more different sectors coupled to each other: The simplest universality class is amended by additional relevant directions, at least one for each additional direction that is added in field space \cite{Eichhorn:2014asa,Eichhorn:2015woa}.

\section{Stability of the potential at large $N$ in $d=5$}
\label{sec:stability-large-N}

At large $N$, a consistent truncation that is sufficient to investigate the stability of the potential is given by Eq.~\eqref{eq:usimp}, i.e., $u(\rho,z)= (m_\phi^2+gz)\rho+v(z)$, where no higher-order terms in the invariant $\rho = \phi^2/2$ appear.
Here, we will argue that this truncation captures all contributions to the effective potential at leading order in $1/N$: Additional interactions are either generated at subleading order in $1/N$ or feed back into the fixed-point equations for $u(\rho,z)$ at subleading order in $1/N$.
Essentially, our argument relies on the fact that only loops which do not contain internal $z$ propagators contribute at leading order in $1/N$.
The only loops satisfying this requirement --~given the truncation with the vertices from Eq.~\eqref{eq:usimp}~-- are those which generate pure $z$ self-interactions, i.e., loops that do not have external $\phi$ lines:
A loop with \emph{external} $\phi$ legs necessarily has to be constructed with the $g\, z \, \phi^2$~vertex which forces the $z$ line to be an internal line and consequently produces a diagram subleading in $1/N$.
Thus the ``dangerous'' interactions are the self-interactions in $z$, as these are generated at leading order in $1/N$ and can couple back into the fixed-point equation for $u(\rho,z)$. For instance, momentum-dependent $z$ self-interactions,  of the form $\propto \, (\partial_\mu z)(\partial^{\mu} z) z^n$, are generated from the $g \, z \, \phi^2$-vertex by a closed $\phi$-loop, which contributes at leading order in $1/N$. However, just as in the case for the pure quartic $O(N)$ model, the back-coupling of this term into the fixed-point equation for $u(\rho,z)$ is given by a closed $z$-loop, which is subleading in $1/N$~\cite{Morris:1997xj}. Thus, it is self-consistent to neglect (momentum-dependent) $z$ self-interactions in our analysis of the stability of the potential at leading order in $1/N$.
Note that our 
argument relies on the analyticity of the potential, i.e., a diagrammatic expansion. 

From now on, we will use the sharp cutoff exclusively, for reasons of computational simplicity. As we will see, it allows us to find an explicit solution to the flow equation for the effective potential that is well-defined at all values of the field $z$.
In the large-$N$ limit all diagrams with an internal $z$ propagator are suppressed, thus
the flow equation for the effective potential $v(z)$ becomes, using Eqs.~\eqref{eq:potflow} and \eqref{eq:usimp},
\begin{align} \label{eq:flow-potential}
 \partial_t v( z) = - d v( z) + \frac{d-2+\eta_z}{2}  z v'( z) + N I_{G,i}^d(m_\phi^2+ g  z),
\end{align}
and the flow of the mass of the $\phi$ field reduces to
\begin{align} \label{eq:flow-mass-phi}
 \partial_t m_\phi^2 & = (\eta_\phi - 2) m_\phi^2.
\end{align}
From this we infer that any fixed point should have $m_{\phi,*}^2 = 0$, since Eq~\eqref{eq:etaphi} becomes for large $N$:
\begin{align}
 \eta_\phi = 0.
\end{align}
As can be seen 
from Fig.~\ref{alldiags}, all explicit contributions to $\beta_g$ are also suppressed by $1/N$. Accordingly, Eq.~\eqref{eq:betag} reduces to the simple form
\begin{align} \label{eq:flow-yukawa}
\partial_t g & = \frac{1}{2}(d-6 + \eta_z)g,
\end{align}
from which the fixed-point requirement again shows that
\begin{align}
 \eta_z = 6-d.
\end{align}
The fixed-point value for $g$ can be obtained from Eq.~\eqref{eq:etaz}, which for large $N$ reduces to $\eta_z = \frac{4v_d}{d} N g^2$, and thus $g^2_* = (6-d)d/(4v_d N)$.

\begin{figure}[t!]
 {\centering \includegraphics[width=0.85\columnwidth]{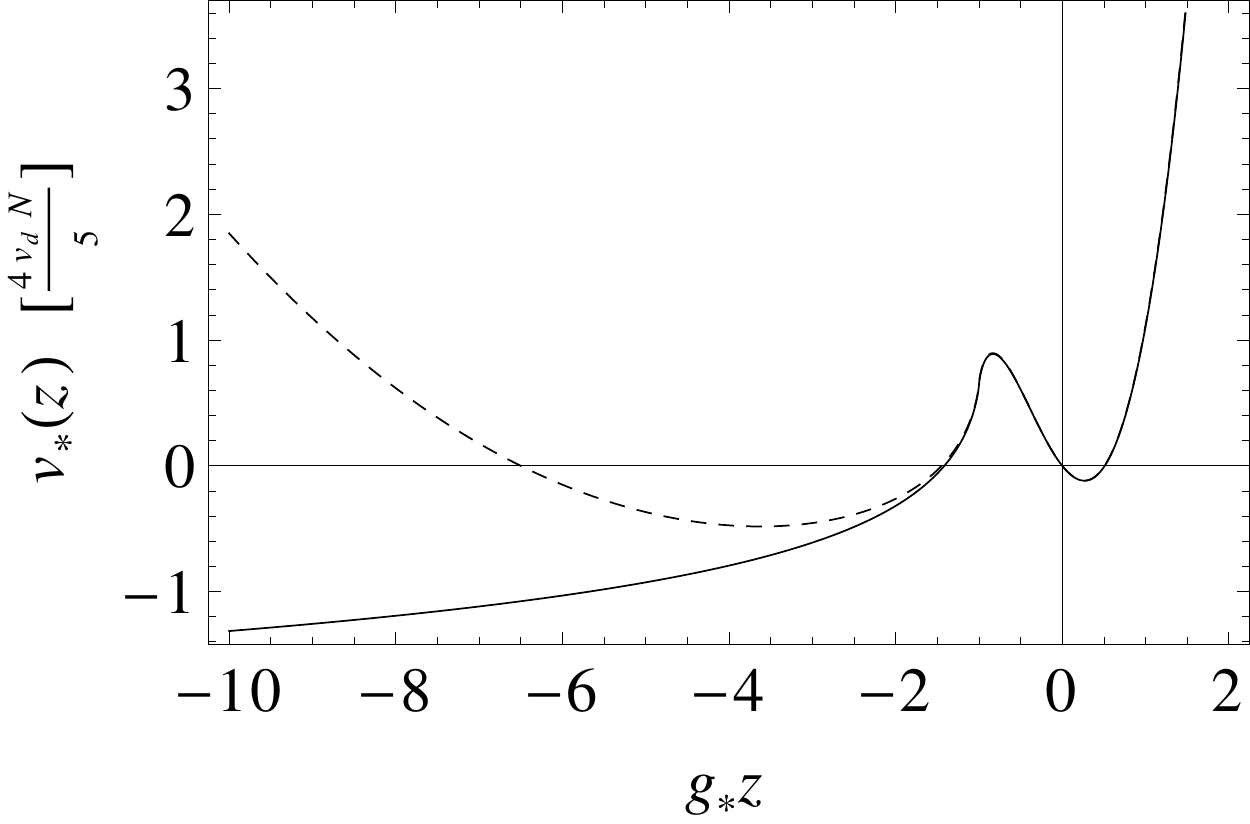}}
 \caption{Large-$N$ fixed-point potential for $c=0$ (solid) and $c=(1/100) \times 4 v_d N / 5$ (dashed).}
 \label{fig:fixed-point-potential}
\end{figure}

The flow of the original $O(N)$ model with quartic coupling $\frac{\mu}{8} \phi^4$ 
can in fact be recovered from Eqs.~\eqref{eq:flow-yukawa} and~\eqref{eq:flow-potential}:
\begin{align}
 \partial_t \left( -\frac{g^2}{m_z^2} \right) & = (d-4) \left( -\frac{g^2}{m_z^2} \right) + 2N v_d \left( -\frac{g^2}{m_z^2} \right)^2,
\end{align}
which is the large-$N$ one-loop beta function for the standard $\frac{\mu}{8}\phi^4$
theory~\cite{herbut2007} upon identification $\mu \equiv -g^2/m_z^2$, which is precisely the relation expected from the Hubbard Stratonovich transformation.
\begin{widetext}

With the known fixed-point values for $g_*$ and $m_{\phi,*}$ we can solve Eq.~\eqref{eq:flow-potential} for the fixed-point potential $v_*(z)$ with $\partial_t v_*(z) \equiv 0$:
\begin{align}\label{eq:fixpot}
v_*( z) = c |g_*  z|^{5/2} + \frac{4v_d N}{5}
\begin{cases}
 (g_* z)^2 - \frac{1}{3} g_*  z + (g_* z)^{5/2} \arctan\left(\sqrt{g_*  z}\right) - \frac{1}{4} \ln(1+g_*  z)^2, & \text{for } 0 \leq g_* z , \\
 (g_* z)^2 -\frac{1}{3} g_*  z-(-g_* z)^{5/2} \text{artanh}\left(\sqrt{-g_*  z}\right) - \frac{1}{4} \ln(1+g_*  z)^2, & \text{for } -1 <g_*  z < 0, \\
  (g_* z)^2 -\frac{1}{3} g_*  z- (-g_* z)^{5/2} \text{artanh}\left(1/\sqrt{-g_*  z}\right) - \frac{1}{4} \ln(1+g_*  z)^2, & \text{for }g_*  z < -1,
\end{cases}
\end{align}
with arbitrary constant $c \in \mathbbm R$.
The form of the potential is plotted for two values of $c$ in Fig.~\ref{fig:fixed-point-potential}.
The potential has the following behavior for small $|g  z| \ll 1$:
\begin{align}
 v_*(z) = c |g_*  z|^{5/2} + 4v_dN\left( - \frac{1}{6} g_*  z + \frac{1}{4} (g_*  z)^2 + \frac{1}{3} (g_*  z)^3\right) + \mathcal O((g_*  z)^4),
\end{align}
while for large $|g_*  z| \gg 1$ we find:
\begin{align}
v_*( z) = c |g_*  z|^{5/2}  + \frac{4v_dN}{5} \left[ \frac{\pi}{2} \Theta(g_*  z)|g_*  z|^{5/2} - \frac{1}{4} \ln((g_*  z)^2) - \frac{1}{5} \right] + \mathcal O(1/|g_*  z|),
\end{align}
with $\Theta(x)$ being the Heaviside step function.
\end{widetext}
As discussed, e.g., in \cite{Morris:1994ki,Morris:1994jc,Morris:1997km}, the solution at large values of $z$ is given by the dimensional scaling (including the anomalous dimension) 
which in our case gives
$v_{\ast}(z) =A_+\, z^{5/2}$ at positive $z$. If the fixed-point potential is real, the corresponding expression for negative $z$ should be $v_{\ast}(z) =A_{-}\, |z|^{5/2}$. 
Focussing on the case $c=0$, the large-$z$ scaling is produced by the same terms that admit a Taylor expansion at low $z$, i.e., the large $z$-scaling provides a boundary condition for the analytic solution. 
This case is clearly realized for $z>0$, and provides $A_+= \frac{4\pi}{10}v_d N\, g_{\ast}$. On the other hand, we find $A_{-}=0$. This is a special case of the general scaling ansatz, and shows that $A_{+}\neq A_{-}$  for our case.

Note a major difference of our model to the case of, e.g., the Wilson-Fisher fixed point with one field species: 
In the latter case, the solution to the homogenous equation, i.e., the one determined by dimensional scaling, only holds at large field values, as the threshold function depends on the derivatives of the potential. In that case, the right-hand side of the Wetterich equation contains a term $\sim 1/(1+v'')$. Only in the asymptotic large-field regime can this term be neglected, and the dimensional scaling ansatz therefore only describes the behavior of the solution \emph{at large fields}, but not everywhere.
In contrast, our case features a threshold function that is independent of the potential $v(z)$, as the fluctuating modes dominating at large $N$ are not those of the $z$-field. 
Thus, the solution to the homogenous equation, $c\, |g_{\ast}z|^{5/2}$ can be \emph{added} to that of the inhomogeneous equation at all field values and $c$ can be chosen freely. 
It is only restricted by analyticity-considerations, but any value of $c$ provides a solution to the fixed-point equation. We can now ask whether this property will persist at finite $N$. 
In that case, the $z$-fluctuations will enter the right-hand-side of the equation and an additional term $\sim 1/(1+v'')$ will appear. Then, the situation will be analogous to that of the Wilson-Fisher fixed point with one field species: The dimensional scaling ansatz $v_{\ast} \sim |z|^{5/2}$ will only hold asymptotically, at large field values. It will \emph{not} be possible to add a term $|z|^{5/2}$ to the solution. Thus, the case of finite $N$ will restrict the viable solutions to $c=0$. 
Demanding a continuous large-$N$ limit thus suggests to set $c=0$ everywhere.

If we insist on the large-$N$ fixed-point potential being analytic in $z$, we must choose $c = 0$. 
In this case the potential is unbounded from below. 
On the other hand, if we allow a small non-analyticity, e.g., with $0< c \ll 1$, the fixed-point potential becomes completely stable (dashed line in Fig.~\ref{fig:fixed-point-potential}).
Within perturbation theory, a non-analytic contribution to the microscopic action would prohibit a straightforward interpretation in terms of Feynman diagrams. 
Even beyond perturbation theory, one would usually demand a theory space that can be spanned in terms of functions that admit a Taylor expansion. 
Admitting further non-analytic functions to extend the basis in theory space presumably could result in problems with predictivity, as each of these functions comes with a coupling that could become relevant. As a simple example, consider all functions of the form $\frac{1}{{z}^{\alpha}}$, with $\alpha\in \mathbb{R}_{+}$. These come with couplings of increasing canonical mass dimension, and therefore result in an infinite number of free parameters, if the theory is considered in the vicinity of the Gaussian fixed point. 
For these reasons, we conclude that 
$c=0$ seems to be the only viable choice in a perturbative context\footnote{On the other hand, explicit examples are known where fixed-point potentials exhibit nonanalyticities, which are nonperturbative phenomena, see, e.g., \cite{David:1984we}.}.
We thus conclude that the potential is not stable in the large $N$ limit in the $z$ direction, at least if we insist on analyticity. 
Our results indicate that the situation for the $O(N)$ model with an additional scalar coupled through a cubic interaction is analogous to that for the pure quartic $O(N)$ model \cite{Percacci:2014tfa}: There, also no analytic fixed-point solution that features a stable fixed-point potential exists at large $N$. 
As a next step, we will investigate the $\phi_i$ direction, and also consider the situation at finite $N$.

\section{Potential stability at finite $N$}
\label{sec:twofield}

\begin{figure}[t!]
 {\centering \includegraphics[width=0.85\columnwidth]{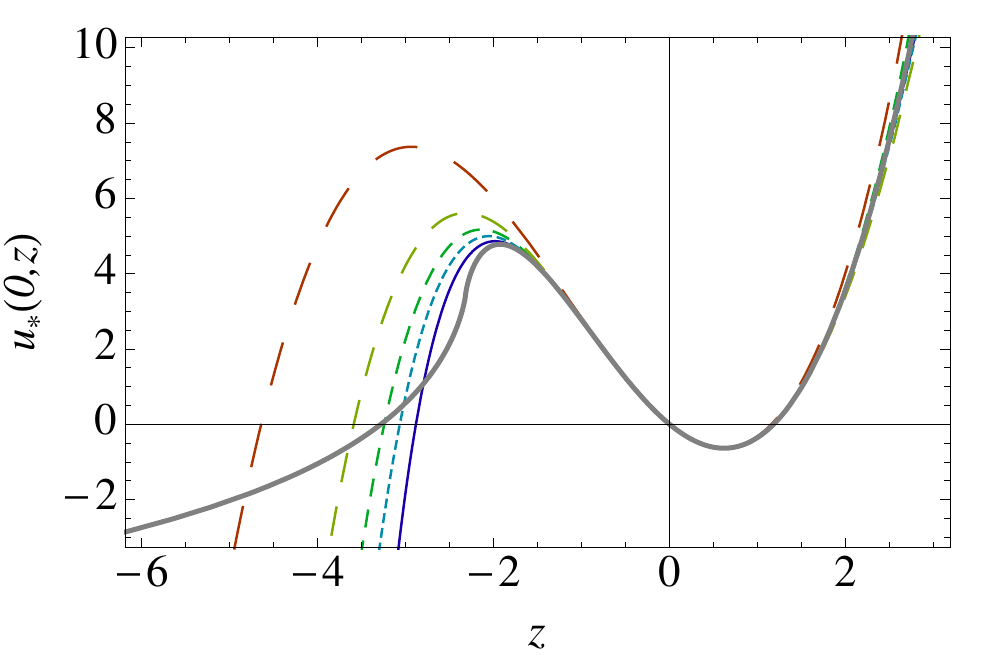}}
 \caption{Comparison of exact large-$N$ fixed-point solution (thick gray, $c=0$) with truncated LPA results at $N=10\,000$ in the sharp cutoff scheme. From long to shorter dashing: LPA3${}^\prime$, LPA4${}^\prime$, LPA5${}^\prime$, LPA6${}^\prime$, LPA8${}^\prime$. Close to the local minimum the LPA perfectly captures the fixed-point potential.}
 \label{fig:fixed-point-potential-LPA}
\end{figure}

\begin{figure}[b!]
\includegraphics[width=.7\columnwidth]{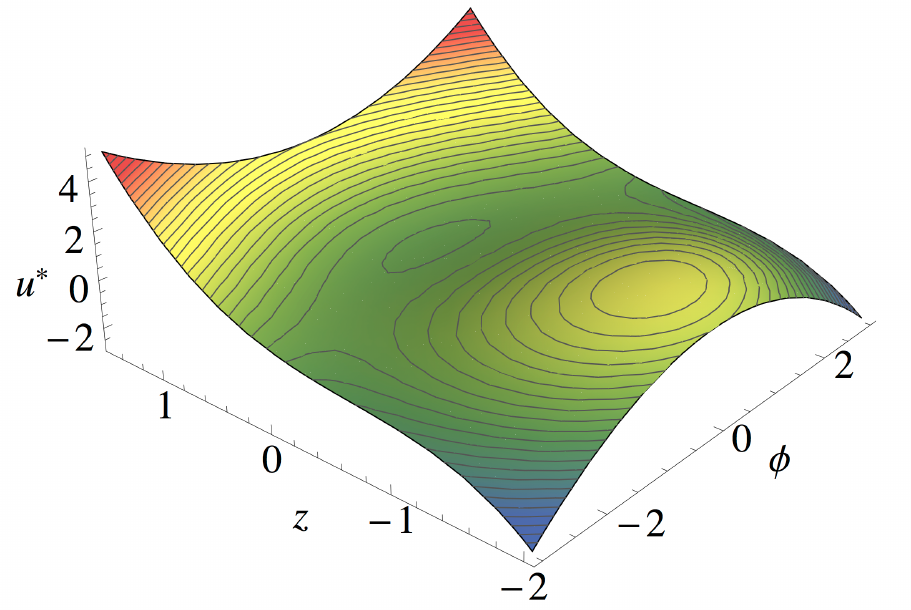}
\includegraphics[width=.7\columnwidth]{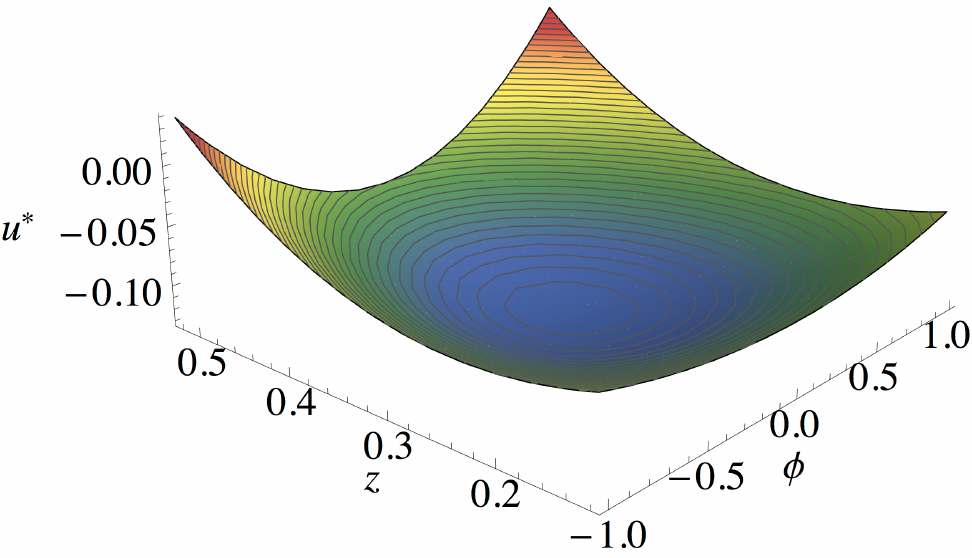}
\caption{Fixed point potential at $\epsilon=1$ and $N=10\,000$. The lower panel clearly exhibits the local minimum of the 
potential. We use an LPA${}^\prime$ with maximally $\phi^8$ and $z^4$.}
\label{fig:potential01-a}
\end{figure}

We are interested in the properties of the potential in the vicinity of $z=0=\phi$, in the two-dimensional field space spanned by $z$ and $\phi$.
This can efficiently be captured by considering a two-dimensional Taylor expansion in the bosonic fields $\rho$ and $z$
\begin{align}\label{eq:SYM2}
	u(\rho,z)=m_\phi^2 \rho+\lambda_1 z + \sum_{n+m\geq2}\frac{\lambda_{n,m}}{n!m!}z^n\rho^m\,.
\end{align}
Then, we have $\lambda_{2,0}=m_z^2$, $\lambda_{1,1}=g$ and $\lambda_{3,0}=\lambda$, cf. Eq.~\eqref{eq:simplesttrunc}.

To highlight that such an expansion already works very well on a quantitative level at low orders of the expansion in the vicinity of the local minimum, we compare the results in the LPA${}^\prime$ at different orders to the full solution at large $N$, cf.~Fig.~\ref{fig:fixed-point-potential-LPA}. On the other hand, just as one should expect within a local expansion, the global properties are captured less well; in particular the LPA${}^\prime$ overestimates the drop of the potential at negative ${z}$.

\begin{figure}[top!]
\includegraphics[width=0.8\columnwidth]{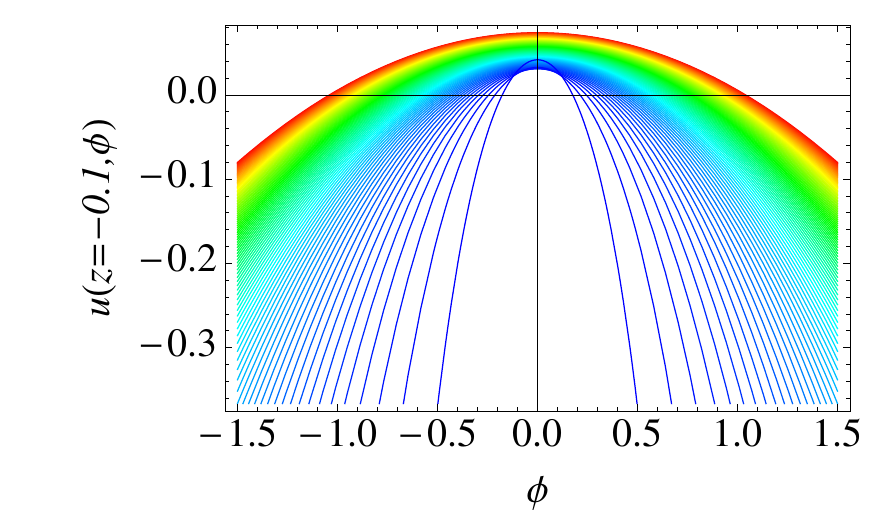}
\includegraphics[width=0.8\columnwidth]{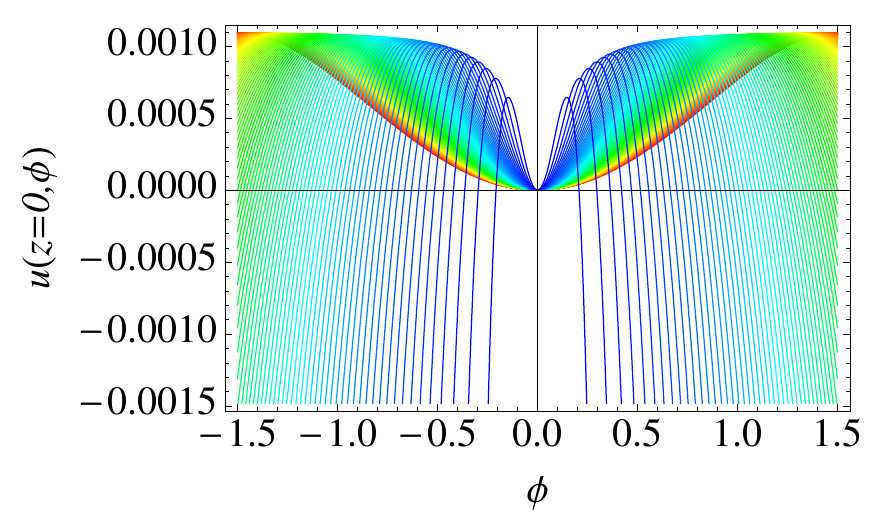}
\includegraphics[width=0.8\columnwidth]{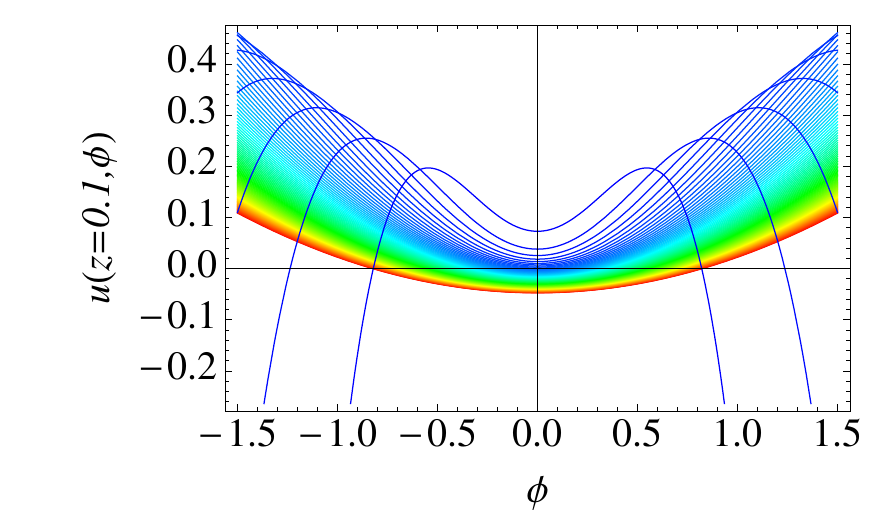}
\caption{Fixed point potential in the sharp cutoff scheme at $\epsilon=1$ and $N=1000-10\,i, i\in \{ 1, 2,...,99\}$ (from red to blue) plotted along the $\phi$ axis at $z\in \{-0.1,0,0.1\}$ (from left to right). The LPA${}^\prime$ takes into account up to $z^4$ and $\phi^4$.}
\label{fig:potential01-N}
\end{figure}
\begin{figure}[top!]
\includegraphics[width=0.8\columnwidth]{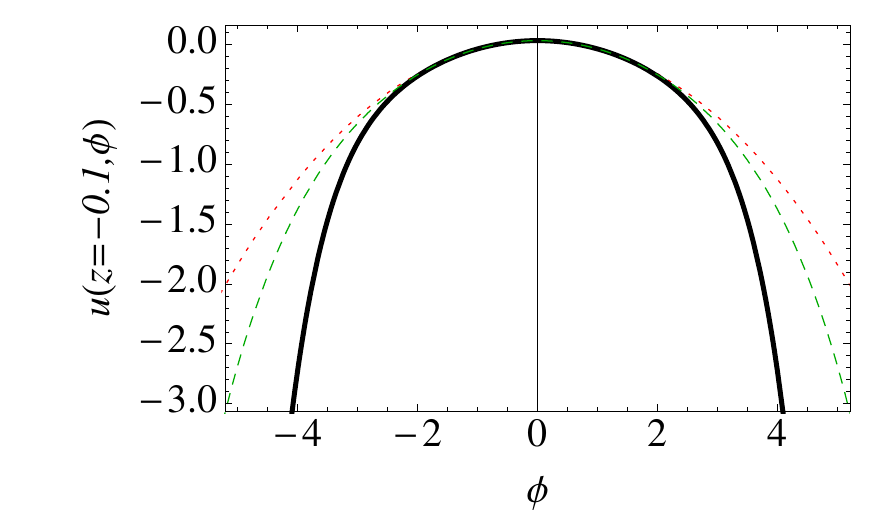}
\includegraphics[width=0.8\columnwidth]{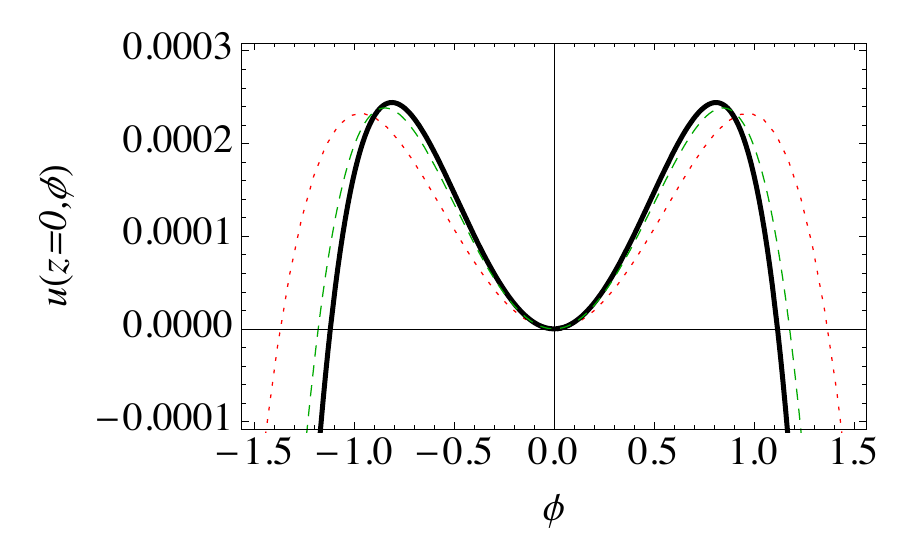}
\includegraphics[width=0.8\columnwidth]{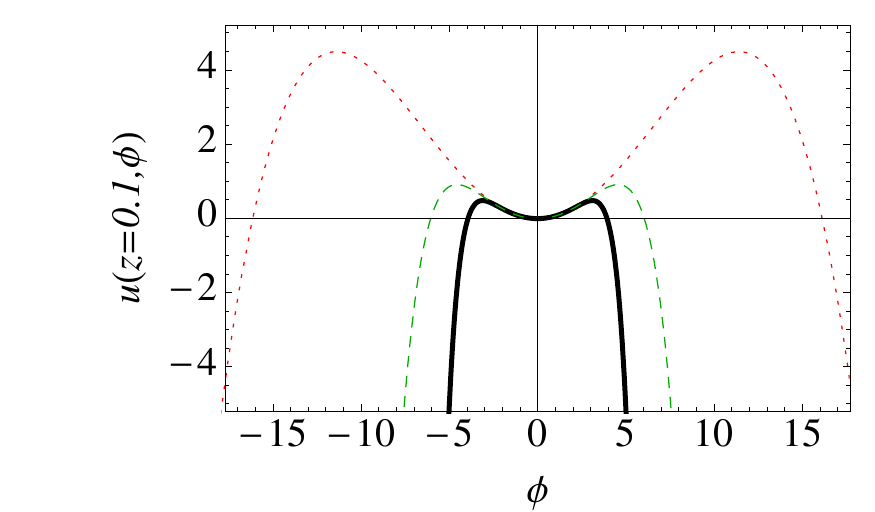}
\caption{Convergence of the two-field potential within different orders of the LPA in $\phi$: LPA4${}^\prime$ (red dotted), LPA6${}^\prime$ (green dashed), LPA8${}^\prime$ (black solid) at $\epsilon=1$ and $N=1000$.}
\label{fig:potential02-N}
\end{figure}

We now use the two-field expansion, Eq.~\eqref{eq:SYM2}, to investigate the fixed point at finite $N$. We observe the existence of a local minimum in the two-dimensional space spanned by $\phi$ and $z$, cf.~Fig.~\ref{fig:potential01-a}. In the $\phi$ direction, the local minimum is generated by a positive mass term, $m_{\phi}^2>0$. All higher-order terms in the $\phi$ direction are negative, $\lambda_{0,m}<0$ for $m>1$. This results in an instability of the potential towards larger absolute values of $\phi$. While the expansion of a globally stable potential can feature negative coefficients at some order in the fields, the coefficients typically alternate in sign in the case of a globally stable potential. In contrast, we observe that all coefficients $\lambda_{0,m}<0$ for $m>1$ are negative, suggesting the onset of instability.

As a next step, we explore whether quantum fluctuations of $z$, which become more important at smaller $N$ feature signatures of a stabilization mechanism for the fixed-point potential: In fact, our result indicates the opposite behavior as expected from our analysis of the constant $c$ in Eq.~\eqref{eq:fixpot}: Towards negative $z$, we observe an unstable potential. Moreover, the lower the value of $N$ the faster the onset of instability towards large $\phi$, see~Fig.~\ref{fig:potential01-N}. As both mass terms, $m_z^2$ and $m_\phi^2$, stay positive at the fixed point and a local minimum always exists at $\phi=0$. Our results however do not indicate a stabilization of that local minimum, as all higher-order couplings stay negative, see Fig.~\ref{fig:potential02-N} for a study of the convergence within the LPA${}^\prime$. In the region where the LPA${}^\prime$ converges around the local minimum, our results indicate that the minimum cannot be the global one. Let us emphasize that a local expansion of the flow equation for the fixed-point potential cannot efficiently detect the formation of a second, global minimum. If such a second minimum existed, and the potential were bounded from below, the situation would correspond to metastability. To explore this potential possibility, global-solution methods must be applied, such as, e.g., those of Ref.~\cite{Borchardt:2015rxa}. With the local expansion employed here we can only exclude that the local minimum close to the origin $\phi=z=0$ appears \emph{not} to be a global minimum as our local expansion captures the onset of instability.

The fixed-point potential is unstable in the $z$ direction at large $N$, unless we allow for the existence of non-analytic terms which we conjecture to also violate continuity of the solution in $N$, and we do not detect a stabilization mechanism at finite $N$. We thus conclude that our results suggest that the fixed-point potential is unstable, at least in the vicinity of the origin.

\section{Conclusions}
\label{sec:conclusions}

We have explored an $O(N)$ symmetric model with an $O(N)$ vector $\phi$ coupled to an additional composite field $z$ via a cubic interaction. It has recently been proposed that this model features an UV fixed point in $d=5$ dimensions~\cite{Fei:2014yja}, thus representing another candidate for an asymptotically safe quantum field theory. Within a one-loop and higher-order $\epsilon$ expansion in dimension $d=6-\epsilon$, the fixed point was found to 
lie at
real values of the couplings only above a critical number of components $N_c$ of the vector field $\phi$. 
The estimate for $N_c$ when extrapolated to the relevant case $\epsilon = 1$ has been found to drastically depend on the order of the $\epsilon$ expansion. Thus, a central question is whether one can find an estimate for $N_c$ beyond the perturbative expansion. For our study, we have employed the functional renormalization group, which can
capture the relevant physics in any dimension directly. In the vicinity of $d=6$, our method reproduces the results for the fixed-point values and anomalous dimensions found within the $\epsilon$ expansion.

In the introduction, we have identified three central questions, which we have set out to answer.
\begin{itemize}
	\item[1.] What is the true  $N_c$ in five dimensions?
\end{itemize}
 Following this fixed point to $d<6$, we find that $N_c$ decreases rapidly, in qualitative but not quantitative agreement with the $\epsilon$-expansion results. Within our approximation, we find that a stable and real fixed point exists all the way down to $N=1$ for all dimensions $4<d < d_c$ with $d_c \approx 5.65$, cf.\ Fig.~\ref{Ndexistence}. In this context, stability refers to the number of relevant directions, which we find to be three, see below.
 Our results thus suggest that there exists no $N_c$ for all $N \geq 1$ in $d=5$.

\begin{itemize}	
	\item[2.] Does the classical equivalence of the cubic theory defined by Eq.~\eqref{eq:lagrangian01} and the original quartic $O(N)$ model continue on the quantum level?
\end{itemize}
We have explored the universal scaling exponents of the cubic model, explicitly comparing with those of the pure $O(N)$ universality class at large $N$. We find that two of the relevant critical exponents precisely agree, while the cubic model features an additional RG relevant direction with positive critical exponent. Thus, the universality class of the pure quartic $O(N)$ model is actually embedded in that of the cubic model.  Our results imply that the models are not fully equivalent, unless one already restricts some of the couplings to lie on the critical hypersurface of the fixed point. This suggests that physically, the additional field $z$ is in fact a new, independent degree of freedom in the model, instead of simply capturing momentum-dependent interactions in the pure $O(N)$ model in a more efficient way.

\begin{itemize}
	\item[3.] Does the cubic model feature a stable fixed-point potential?
\end{itemize}
We have explored the global properties of the fixed-point potential at large $N$, which turned out to be unstable towards negative $z$, unless nonanalytic terms are allowed in the fixed-point action.
This is similar to the results of the analysis of the $O(N)$ model within the original quartic formulation~\cite{Percacci:2014tfa}.

In order to address the situation at finite $N$, we have evaluated the fixed-point potential in the two-dimensional field space spanned by $\phi$ and $z$. 
While we uncover the existence of a local minimum, it only exists due to a positive mass term for $\phi$, while all higher-order couplings in $\phi$ are negative. 
This suggests the onset of instability for larger values of $\phi$. Moreover, our local expansion features an instability towards large negative $z$, as is already implied by the analytic solution for large $N$.
We should emphasize that our finite-$N$ study, based on a local expansion of the potential, cannot firmly exclude the possibility that quantum fluctuations generate a second minimum at larger field values. 
This could result in a potential that is globally bounded from below, and would imply that the minimum that we have investigated here is in fact metastable. 
To address this question, global methods such as those in \cite{Borchardt:2015rxa} should be employed. 
This is left for a future study. Here, we can only conclude that quantum fluctuations do not appear to stabilize the local minimum, but in fact trigger a faster onset of its instability towards lower $N$. 

The fixed point of the cubic model hence appears to suffer from the same deficiency as the Wilson-Fisher fixed point of the original quartic $O(N)$ model in $d>4$, at least at large $N$ and as long as one insists on an analytic fixed-point potential. 
This raises the question whether it may at all be possible, at least in principle, to find an explicit (e.g., lattice) construction with positive measure whose continuum limit realizes our fixed point. 
The known lattice $O(N)$ models \cite{Aizenman:1981du, Frohlich:1982tw, Wolff:2009ke} exhibit (in agreement with the perturbative expectation \cite{Wilson:1973jj}) trivial critical behavior, governed by the noninteracting Gaussian fixed point. It is not clear whether a lattice study can assess the nontrivial fixed point of our cubic theory, as it would require to overcome the notorious difficulties that arise through the nonanalyticities of the fixed-point potential.

In summary, we have found an interacting fixed point of the cubic $O(N)$ model which exists for all $N$ in $d=5$. The universality class it defines includes the standard $O(N)$ universality class (analytically continued to $d>4$), but features an additional RG relevant direction. On the quantum level, the cubic $O(N)$ model hence is not fully equivalent to the original quartic $O(N)$ model.
Our study indicates that no stable analytic fixed-point potential can be found, questioning the 
realization of the fixed point from underlying a well-defined quantum field theory.

\begin{acknowledgments}
We thank Holger Gies and Daniel F.~Litim for helpful discussions and comments on this manuscript. We also thank Igor Boettcher and Jan M.~Pawlowski for discussions and Igor F.~Herbut for collaboration on an earlier related project.
M.M.S.~thanks the Imperial College for hospitality.  M.M.S. is
supported by Grant No. ERC-AdG-290623 and DFG Grant No. SCHE 1855/1-1. L.J.\ acknowledges support by the DFG under Grant No.\ SFB1143. The work of A.E. is supported by an Imperial
College Junior Research Fellowship.
\end{acknowledgments}



\thebibliography{99}

\bibitem{Fei:2014yja} 
  L.~Fei, S.~Giombi and I.~R.~Klebanov,
  Phys.\ Rev.\ D {\bf 90}, 025018 (2014)
  [arXiv:1404.1094 [hep-th]].

\bibitem{Fei:2014xta} 
  L.~Fei, S.~Giombi, I.~R.~Klebanov and G.~Tarnopolsky,
  Phys.\ Rev.\ D {\bf 91}, 045011 (2015)
  [arXiv:1411.1099 [hep-th]].
  
\bibitem{Weinberg:1980gg}
  S.~Weinberg,
  in {\it General Relativity}, edited by S. Hawking and W. Israel (Cambridge University Press, Cambridge, England, 1979), pp. 790-831.
  
\bibitem{Gies:2013pma}
  H.~Gies, S.~Rechenberger, M.~M.~Scherer and L.~Zambelli,
  Eur.\ Phys.\ J.\ C {\bf 73} (2013) 2652
  [arXiv:1306.6508 [hep-th]].

\bibitem{Litim:2014uca} 
  D.~F.~Litim and F.~Sannino,
  JHEP {\bf 1412}, 178 (2014)
  [arXiv:1406.2337 [hep-th]].

\bibitem{Reuter:1996cp} 
  M.~Reuter,
  Phys.\ Rev.\ D {\bf 57}, 971 (1998)
  [hep-th/9605030].
  
\bibitem{Codello:2008vh} 
  A.~Codello, R.~Percacci and C.~Rahmede,
  Annals Phys.\  {\bf 324}, 414 (2009)
  [arXiv:0805.2909 [hep-th]].
  
%
\bibitem{Benedetti:2009rx} 
  D.~Benedetti, P.~F.~Machado and F.~Saueressig,
  Mod.\ Phys.\ Lett.\ A {\bf 24}, 2233 (2009)
  [arXiv:0901.2984 [hep-th]].
  
\bibitem{Falls:2013bv} 
  K.~Falls, D.~F.~Litim, K.~Nikolakopoulos and C.~Rahmede,
  arXiv:1301.4191 [hep-th].
  
 \bibitem{Becker:2014qya} 
  D.~Becker and M.~Reuter,
  Annals Phys.\  {\bf 350}, 225 (2014)
  [arXiv:1404.4537 [hep-th]].
  
\bibitem{Dona:2013qba} 
  P.~Don\`a, A.~Eichhorn and R.~Percacci,
 Phys.\ Rev.\ D {\bf 89}, 084035 (2014)
  [arXiv:1311.2898 [hep-th]].
  
\bibitem{Fradkin:1987ks} 
  E.~S.~Fradkin and M.~A.~Vasiliev,
  Phys.\ Lett.\ B {\bf 189}, 89 (1987).
  
\bibitem{Vasiliev:1990en} 
  M.~A.~Vasiliev,
  Phys.\ Lett.\ B {\bf 243}, 378 (1990).
  
\bibitem{Vasiliev:1992av} 
  M.~A.~Vasiliev,
  Phys.\ Lett.\ B {\bf 285}, 225 (1992).
  
\bibitem{Vasiliev:1995dn} 
  M.~A.~Vasiliev,
  Int.\ J.\ Mod.\ Phys.\ D {\bf 5}, 763 (1996)
  [hep-th/9611024].
  
\bibitem{Vasiliev:1999ba} 
  M.~A.~Vasiliev,
  In {\it The many faces of the superworld}, edited by M. A. Shifman (World Scientific, Singapore, 2000), pp. 33-610
  [hep-th/9910096].
  
\bibitem{Vasiliev:2003ev} 
  M.~A.~Vasiliev,
  Phys.\ Lett.\ B {\bf 567}, 139 (2003)
  [hep-th/0304049].
  
    %
\bibitem{Klebanov:2002ja} 
  I.~R.~Klebanov and A.~M.~Polyakov,
  Phys.\ Lett.\ B {\bf 550}, 213 (2002)
  [hep-th/0210114].
  
\bibitem{Giombi:2014iua} 
  S.~Giombi, I.~R.~Klebanov and B.~R.~Safdi,
  Phys.\ Rev.\ D {\bf 89}, 084004 (2014)
  [arXiv:1401.0825 [hep-th]].
  
\bibitem{Gies:2009hq} 
  H.~Gies and M.~M.~Scherer,
  Eur.\ Phys.\ J.\ C {\bf 66}, 387 (2010)
  [arXiv:0901.2459 [hep-th]].
  
\bibitem{Gies:2009sv}
  H.~Gies, S.~Rechenberger and M.~M.~Scherer,
  Eur.\ Phys.\ J.\ C {\bf 66} (2010) 403
  [arXiv:0907.0327 [hep-th]].
  
\bibitem{Scherer:2009wu}
  M.~M.~Scherer, H.~Gies and S.~Rechenberger,
  Acta Phys.\ Polon.\ Supp.\  {\bf 2}, 541 (2009)
  [arXiv:0910.0395 [hep-th]].

\bibitem{Vacca:2015nta}
  G.~P.~Vacca and L.~Zambelli,
  Phys.\ Rev.\ D {\bf 91}, 125003 (2015)
  [arXiv:1503.09136 [hep-th]].
  
\bibitem{Gracey:2015tta} 
  J.~A.~Gracey,
  Phys.\ Rev.\ D {\bf 92}, 025012 (2015)
  [arXiv:1506.03357 [hep-th]].

\bibitem{Nakayama:2014yia} 
  Y.~Nakayama and T.~Ohtsuki,
  Phys.\ Lett.\ B {\bf 734}, 193 (2014)
  [arXiv:1404.5201 [hep-th]].

\bibitem{Chester:2014gqa} 
  S.~M.~Chester, S.~S.~Pufu and R.~Yacoby,
  Phys.\ Rev.\ D {\bf 91}, 086014 (2015)
  [arXiv:1412.7746 [hep-th]].
 
\bibitem{Bae:2014hia} 
  J.~B.~Bae and S.~J.~Rey,
  arXiv:1412.6549 [hep-th].

\bibitem{Herbut:2015zqa} 
  I.~F.~Herbut and L.~Janssen,
  Phys. Rev. D {\bf 93}, 085005 (2016)
  [arXiv:1510.05691 [hep-th]].
  
\bibitem{coleman1973}
  S.~R.~Coleman and E.~J.~Weinberg,
  Phys.\ Rev.\ D {\bf 7}, 1888 (1973).

\bibitem{halperin1974} 
  B.~I.~Halperin, T.~C.~Lubensky and S.~K.~Ma,
  Phys.\ Rev.\ Lett.\  {\bf 32}, 292 (1974).

\bibitem{kolnberger1990}
S. Kolnberger and R. Folk, 
Phys. Rev. B {\bf 41}, 4083 (1990).

\bibitem{Bergerhoff:1995zm} 
  B.~Bergerhoff, D.~Litim, S.~Lola and C.~Wetterich,
  Int.\ J.\ Mod.\ Phys.\ A {\bf 11}, 4273 (1996)
  [cond-mat/9502039].

\bibitem{herbut1997}
  I. F. Herbut and Z. Tesanovic, 
  Phys. Rev. Lett. {\bf 78}, 980 (1997)
  [cond-mat/9605185].

\bibitem{herbut1996}
I. F. Herbut and Z. Tesanovic, Phys. Rev. Lett. {\bf 76}, 4588 (1996).

\bibitem{bergerhoff1996}
B. Bergerhoff, F. Freire, D. F. Litim, S. Lola, and C. Wetterich, 
Phys. Rev. B {\bf 53}, 5734 (1996).

\bibitem{dasgupta1981}
C. Dasgupta and B. I. Halperin, Phys. Rev. Lett. {\bf 47}, 1556 (1981);
%
J. Bartholomew, Phys. Rev. B {\bf 28}, 5378(R) (1983);
%
S. Mo, J. Hove, and A. Sudbo, Phys. Rev. B {\bf 65}, 104501 (2002).
 
\bibitem{Freire:2000sx} 
  F.~Freire and D.~F.~Litim,
  Phys.\ Rev.\ D {\bf 64}, 045014 (2001)
  [hep-ph/0002153].

\bibitem{herbut2007}
I.~Herbut, \textit{A modern approach to critical phenomena},
Cambridge University Press, Cambridge, 2007.   

\bibitem{Berges:2000ew}
  J.~Berges, N.~Tetradis and C.~Wetterich,
  Phys.\ Rept.\  {\bf 363} (2002) 223
  [hep-ph/0005122].

\bibitem{Percacci:2014tfa} 
  R.~Percacci and G.~P.~Vacca,
  Phys.\ Rev.\ D {\bf 90}, 107702 (2014)
  [arXiv:1405.6622 [hep-th]].
  
\bibitem{Wetterich:1992yh} 
  C.~Wetterich,
  Phys.\ Lett.\ B {\bf 301}, 90 (1993).
  
\bibitem{epsilongravity}
  S.~M.~Christensen and M.~J.~Duff,
  Phys.\ Lett.\ B {\bf 79}, 213 (1978);
  R.~Gastmans, R.~Kallosh and C.~Truffin,
  Nucl.\ Phys.\ B {\bf 133}, 417 (1978);
  H.~Kawai and M.~Ninomiya,
  Nucl.\ Phys.\ B {\bf 336}, 115 (1990);
  H.~Kawai, Y.~Kitazawa and M.~Ninomiya,
  Nucl.\ Phys.\ B {\bf 393}, 280 (1993)
  [hep-th/9206081];
  {\it ibid.} {\bf 467}, 313 (1996)
  [hep-th/9511217].
  
\bibitem{Morris:1993qb} 
  T.~R.~Morris,
  Int.\ J.\ Mod.\ Phys.\ A {\bf 9}, 2411 (1994)
  [hep-ph/9308265].
  
\bibitem{Ellwanger:1993mw} 
  U.~Ellwanger,
  Z.\ Phys.\ C {\bf 62}, 503 (1994)
  [hep-ph/9308260].

\bibitem{Polonyi:2001se}
  J.~Polonyi,
  Central Eur.\ J.\ Phys.\  {\bf 1}, 1 (2003) 
 [hep-th/0110026].
%
\bibitem{Pawlowski:2005xe}
  J.~M.~Pawlowski,
  Annals Phys.\  {\bf 322} (2007) 2831 
  [arXiv:hep-th/0512261].
%

\bibitem{Gies:2006wv}
  H.~Gies, Lect.\ Notes Phys.\ {\bf 852}, 287 (2012)
  [arXiv:hep-ph/0611146]. 

\bibitem{Delamotte:2007pf} 
  B.~Delamotte,
  Lect.\ Notes Phys.\  {\bf 852}, 49 (2012)
  [cond-mat/0702365].
  
\bibitem{Rosten:2010vm}
  O.~J.~Rosten,
  arXiv:1003.1366 [hep-th].
  
\bibitem{Braun:2011pp} 
  J.~Braun,
  J.\ Phys.\ G {\bf 39}, 033001 (2012)
  [arXiv:1108.4449 [hep-ph]].

\bibitem{metzner2011} 
W.~Metzner et al., Rev. Mod. Phys. {\bf 84}, 299 (2012).

\bibitem{litim01}  
  D.~F.~Litim,
  Phys.\ Rev.\ D {\bf 64}, 105007 (2001)
  [hep-th/0103195].
  
\bibitem{Braun:2014wja} 
  J.~Braun, H.~Gies, L.~Janssen and D.~Roscher,
  Phys.\ Rev.\ D {\bf 90}, 036002 (2014)
  [arXiv:1404.1362 [hep-ph]].

\bibitem{Reuter:2001ag} 
  M.~Reuter and F.~Saueressig,
  Phys.\ Rev.\ D {\bf 65}, 065016 (2002)
  [hep-th/0110054].

\bibitem{Janssen:2012pq} 
  L.~Janssen and H.~Gies,
  Phys.\ Rev.\ D {\bf 86}, 105007 (2012)
  [arXiv:1208.3327 [hep-th]].
 
\bibitem{litim}  
D.~F.~Litim,
  Phys.\ Lett.\ B {\bf 486}, 92 (2000)
  [hep-th/0005245];
%
  Int.\ J.\ Mod.\ Phys.\ A {\bf 16}, 2081 (2001)
  [hep-th/0104221];
%
  Nucl.\ Phys.\ B {\bf 631}, 128 (2002)
  [hep-th/0203006].
    
\bibitem{Pawlowski:2015mlf}
  J.~M.~Pawlowski, M.~M.~Scherer, R.~Schmidt and S.~J.~Wetzel,
  arXiv:1512.03598 [hep-th].
  
\bibitem{Jaeckel:2002rm} 
  J.~Jaeckel and C.~Wetterich,
  Phys.\ Rev.\ D {\bf 68}, 025020 (2003)
  [hep-ph/0207094].
  
\bibitem{Gies:2001nw} 
  H.~Gies and C.~Wetterich,
  Phys.\ Rev.\ D {\bf 65}, 065001 (2002)
  [hep-th/0107221].
  
\bibitem{Floerchinger:2009uf} 
  S.~Floerchinger and C.~Wetterich,
  Phys.\ Lett.\ B {\bf 680}, 371 (2009)
  [arXiv:0905.0915 [hep-th]].
  
\bibitem{Gehring:2015vja} 
  F.~Gehring, H.~Gies and L.~Janssen,
  Phys.\ Rev.\ D {\bf 92}, 085046 (2015)
  [arXiv:1506.07570 [hep-th]].

\bibitem{Mati:2016wjn} 
  P.~Mati,
  arXiv:1601.00450 [hep-th].

\bibitem{Litim:1995ex} 
  D.~Litim and N.~Tetradis,
  hep-th/9501042.

\bibitem{Tetradis:1995br} 
  N.~Tetradis and D.~F.~Litim,
  Nucl.\ Phys.\ B {\bf 464}, 492 (1996)
  [hep-th/9512073].
  
\bibitem{ZinnJustin:1998cp} 
  J.~Zinn-Justin,
  hep-th/9810198.
  
\bibitem{Eichhorn:2013zza} 
  A.~Eichhorn, D.~Mesterh\'azy and M.~M.~Scherer,
  Phys.\ Rev.\ E {\bf 88}, 042141 (2013)
  [arXiv:1306.2952 [cond-mat.stat-mech]].
  
\bibitem{Eichhorn:2014asa} 
  A.~Eichhorn, D.~Mesterh\'azy and M.~M.~Scherer,
  Phys.\ Rev.\ E {\bf 90}, 052129 (2014)
  [arXiv:1407.7442 [cond-mat.stat-mech]].
  
\bibitem{Eichhorn:2015woa} 
  A.~Eichhorn, T.~Helfer, D.~Mesterh\'azy and M.~M.~Scherer,
  Eur.\ Phys.\ J.\ C {\bf 76}, 88 (2016)
  [arXiv:1510.04807 [cond-mat.stat-mech]].

\bibitem{Morris:1997xj} 
  T.~R.~Morris and M.~D.~Turner,
  Nucl.\ Phys.\ B {\bf 509}, 637 (1998)
  [hep-th/9704202].
  
  %
\bibitem{Morris:1994ki} 
  T.~R.~Morris,
  Phys.\ Lett.\ B {\bf 334}, 355 (1994)
  [hep-th/9405190].
  
\bibitem{Morris:1994jc} 
  T.~R.~Morris,
  Phys.\ Lett.\ B {\bf 345}, 139 (1995)
  [hep-th/9410141].
  
\bibitem{Morris:1997km} 
  T.~R.~Morris,
  hep-th/9709100.
  

\bibitem{David:1984we} 
  F.~David, D.~A.~Kessler and H.~Neuberger,
  Phys.\ Rev.\ Lett.\  {\bf 53}, 2071 (1984).

\bibitem{Borchardt:2015rxa} 
  J.~Borchardt and B.~Knorr,
  Phys.\ Rev.\ D {\bf 91}, 105011 (2015)
  [arXiv:1502.07511 [hep-th]].
  
\bibitem{Aizenman:1981du} 
  M.~Aizenman,
  Phys.\ Rev.\ Lett.\  {\bf 47}, 1 (1981).
  
\bibitem{Frohlich:1982tw} 
  J.~Frohlich,
  Nucl.\ Phys.\ B {\bf 200}, 281 (1982).
  
\bibitem{Wolff:2009ke} 
  U.~Wolff,
  Phys.\ Rev.\ D {\bf 79}, 105002 (2009)
  [arXiv:0902.3100 [hep-lat]],
  and references therein.
  
\bibitem{Wilson:1973jj} 
  K.~G.~Wilson and J.~B.~Kogut,
  Phys.\ Rept.\  {\bf 12}, 75 (1974).
\end{document}